\documentclass[twocolumn]{aastex631}

\accepted{Version October 31 2023}
\usepackage{amsmath}
\usepackage[caption=false]{subfig}
\usepackage[shortlabels]{enumitem}

\begin{document}
\nolinenumbers
\title{A study of two periodogram algorithms for improving the detection of small transiting planets}
\correspondingauthor{Eric D. Feigelson}

\author[0000-0002-6646-4225]{Yash Gondhalekar}
\affiliation{Dept. of CSIS, BITS Pilani K.K Birla Goa Campus, Goa, 403726, Goa, India}

\author[0000-0002-5077-6734]{Eric D. Feigelson}
\affiliation{Department of Astronomy \& Astrophysics,
Pennsylvania State University, 525 Davey Laboratory,
University Park, PA 16802, USA}
\affiliation{Center for Astrostatistics, Pennsylvania State University, 525 Davey Laboratory, University Park PA, 16802, USA}
\affiliation{Center for Exoplanets and Habitable Worlds, Pennsylvania State University, 525 Davey Laboratory, University Park PA, 16802, USA}

\author{Gabriel A. Caceres}
\affiliation{EY-Parthenon, 1540 Broadway, New York NY, 10036, USA}

\author{Marco Montalto}
\affiliation{INAF - Osservatorio Astrofisico di Catania, Via S. Sofia 78, I-95123 Catania, Italy}

\author[0000-0002-8458-604X]{Snehanshu Saha}
\affiliation{Dept. of CSIS and APPCAIR, BITS Pilani K.K Birla Goa Campus, Goa, 403726, Goa, India}
\affiliation{Division of Artificial Intelligence Research, HappyMonk AI, Bangalore, 560078, Karnataka, India}

\begin{abstract}
The sensitivities of two periodograms are compared for weak signal planet detection in transit surveys: the widely used Box-Least Squares (BLS) algorithm following light curve detrending and the Transit Comb Filter (TCF) algorithm following autoregressive ARIMA modeling. Small depth transits are injected into light curves with different simulated noise characteristics. Two measures of spectral peak significance are examined: the periodogram signal-to-noise ratio (SNR) and a False Alarm Probability (FAP) based on the generalized extreme value distribution. The relative performance of the BLS and TCF algorithms for small planet detection is examined for a range of light curve characteristics, including orbital period, transit duration, depth, number of transits, and type of noise. We find that the TCF periodogram applied to ARIMA fit residuals with the SNR detection metric is preferred when short-memory autocorrelation is present in the detrended light curve and even when the light curve noise had white Gaussian noise. BLS is more sensitive to small planets only under limited circumstances with the FAP metric. BLS periodogram characteristics are inferior when autocorrelated noise is present due to heteroscedastic noise and false period detection. Application of these methods to TESS light curves with known small exoplanets confirms our simulation results. The study ends with a decision tree that advises transit survey scientists on procedures to detect small planets most efficiently.  The use of ARIMA detrending and TCF periodograms can significantly improve the sensitivity of any transit survey with regularly spaced cadence.  
\end{abstract} 

\keywords{Period search  (1955), Time series analysis (1916), Transits (1711), Exoplanet detection methods (489), Transit photometry (1709)}

\section{Introduction}\label{sec:intro}

\subsection{Difficulties with detecting small transiting planets}
\label{sec:intro_difficulties}

The transits of giant Jovian planets producing periodic $\sim1$\% dips in brightness can be easily seen in photometric light curves produced by space-based observatories such as COROT \citep{corot2008}, Kepler \citep{Kepler2010}, K2 \citep{K2_2014}, TESS \citep{Ricker15}, and likely in the forthcoming PLATO \citep{plato2014} and Roman Space Telescope \citep{Roman2015} missions. However, achieving planned goals to discover suspected large populations of smaller planets has proved challenging. Predictions that several thousand transiting planets will emerge from analysis of TESS data \citep{Barclay18, Kunimoto22} are, at present, overly optimistic\footnote{A similar discrepancy between predicted and achieved transiting samples was noted by \citet{Horne03} from early ground-based transit surveys.}. Problems arise because transits from smaller rocky planets producing $0.01\%-0.1$\% periodic photometric dips are often masked by other sources of photometric variability \citep{Gilliland11}: rotational modulation of starspots \citep{McQuillan14}; microvariability from stochastic stellar magnetic activity \citep{Aigrain04}; contamination by eclipsing binaries blended in the large pixels of wide-field telescopes \citep{Torres11}; instrumental effects involving satellite operations \citep{Vanderburg14}; red noise \citep{Pont06}; and unavoidable detector photon noise. 

The challenge of small planet detection requires solving complicated problems in time series analysis. Several stages of analysis are needed:
\begin{enumerate}
    \item  The light curve is detrended to remove aperiodic or quasi-periodic variations unrelated to strictly periodic planetary orbits. Detrending is typically pursued using nonparametric or semi-parametric methods such as spline fitting, wavelet transforms, or Gaussian Processes regression \citep[e.g.][]{Jenkins02, Gibson2014, Lightkurve18, Hippke2019Detrending, Feinstein19, Montalto20, ForemanMackey21, Guerrero21}. 
    
    \item The detrended light curve is typically searched for transit-shaped periodic dips using the parametric Box-Least Squares (BLS) algorithm developed by \citet{Kovacs02}. For each trial period, a box-shaped signal is fit to the folded light curve for a range of transit durations and phases. A BLS periodogram is constructed using the strongest signal found at each period, and spectral peaks are investigated as possible transit signals. Kovacs et al. show that this procedure is more sensitive to faint box-shaped dips than Fourier periodograms \citep[or, for irregular observation cadences, Lomb-Scargle periodograms;][]{Scargle82} that search for sinusoidal signals and more sensitive than nonparametric periodograms that search for arbitrarily shaped signals such as phase dispersion minimization \citep{Stellingwerf78}.
    
    \item Procedures are applied to cull False Alarms and astronomical False Positives (particularly contaminant eclipsing binaries in telescopes with low-resolution images) from spectral peaks above some threshold. This typically involves a combination of machine learning classification (such as Random Forest or neural network) and human or automated vetting. Vetting procedures for transit planet detection are discussed by \citet{Thompson15}, \citet{Twicken18},  \citet{Hedges21},  \citet{Guerrero21}, \citet{Melton23b}, and others.   
\end{enumerate}
For the Kepler and TESS space missions, these procedures are used by NASA science teams to generate Kepler and TESS Objects of Interest (KOIs and TOIs) that are then passed to ground-based telescopes for further study \citep{Twicken16, Guerrero21}. 

However, these standard procedures for planet detection may have technical deficiencies that other statistical approaches might improve. At least two issues might be considered. First, detrenders based on a function with a kernel or a mother wavelet with constant bandwidth can miss short-memory variations. The main concern is autoregressive variations characteristic of stellar magnetic activity. Autoregressive behaviors occur when future values of a time series depend, at least in part, on current and past values.  Its presence can be easily checked by plotting the autocorrelation function of the detrended light curve to see if short-memory autoregressive behaviors are present. Formal statistical time series diagnostics, such as the Shapiro-Wilk and Ljung-Box tests, can determine if the light curves deviate from Gaussian white noise. This is a significant problem: 36\% of light curves from TESS Full Frame Images (FFIs) show statistically significant short-memory stochastic variability after spline detrending \citep[][their Figure 5]{Melton23a}. 

Second, periodograms have complicated and poorly understood statistical properties that hinder the straightforward identification of significant peaks representing actual periodic behaviors. Even in classical Fourier analysis, the theorems underlying the statistical distribution of periodogram peaks apply only to the unrealistic situation of an infinitely long, uninterrupted, evenly-spaced data stream of Gaussian white noise with a single sinusoidal signal \citep{Percival09}. The Lomb-Scargle periodogram \citep{Lomb1976, Scargle82} was proposed as an extension to the classical periodogram to alleviate issues arising due to irregular data spacing while providing other statistical benefits. 

However, several issues in periodogram analyses still persist. Spurious spectral peaks can arise from periodic instrumental effects (such as data gaps associated with the satellite orbital period or data downloads) for realistic time series arising from space-based missions. Complicated aliases of true signals often appear \citep{VanderPlas18}. Periodograms can exhibit undesirable behaviors even when no interesting signal is present: \citet{Ofir14} shows that the noise properties of BLS periodograms often have trends in value and noise (heteroscedasticity) as a function of the trial period.

The first issue concerning autoregressive variations that may escape removal by detrenders has a clear treatment using low-dimensional parametric autoregressive moving average (ARMA) models. These are commonly combined with a simple nonparametric detrender involving `differencing' (or differentiating) the time series. The resulting ARIMA modeling (also known as Box-Jenkins analysis) and its many extensions have dominated analyses of stochastic time series for the past 50 years in engineering signal processing, econometrics, and other fields. Autoregressive modeling dominates most textbooks; the foundational text by \citet{Box15} has over 50 thousand citations and a Nobel Prize in Economics was awarded for the non-linear GARCH model that introduced stochastic volatility to the simpler linear ARIMA model. 

ARIMA detrending for transiting planet detection was introduced by \citet{Caceres19a} and found to be effective in reducing unwanted light curve variations in most Kepler and TESS light curves \citep{Caceres19b, Melton23a}. The general utility of ARMA-type modeling for astronomical time domain studies is discussed by \citet{Feigelson18}.  

However, the traditional BLS algorithm can not be applied to ARIMA residuals because the differencing operation changes a sequence of box-shaped into a sequence of double spikes. \citet{Caceres19a} developed the Transit Comb Filter\footnote{The TCF can be considered an extension of the Dirac comb or Shah function used in engineering signal processing.} (TCF) as an alternative algorithm to produce periodograms of ARIMA residuals. As the double spike reflects only the ingress and egress, it would seem to have less information and therefore be less sensitive to weak transits than BLS fitting. However, in searching for small planets in the 4-year Kepler data, \citet{Caceres19b} find that the TCF periodogram has low noise and is remarkably sensitive to faint transits corresponding to Earth- and Mars-sized planets. Similarly, \citet[][their Figure 16]{Melton23c} find that the ARIMA-TCF procedure tends to find smaller planets than other pipelines used to generate TOIs.    

In addition to the choice of the periodogram algorithm, the transit scientist must decide on the best measure for the strength of a spectral peak. The most straightforward approach is to locate the period with the highest periodogram power. However, strong peaks can be produced by autocorrelated noise in the light curve or aliasing associated with an irregular cadence. Noting heteroscedasticity and trends in periodograms' response to noise, \citet{Ofir14} recommends using local signal-to-noise ratios of the detrended BLS periodogram. 

Evaluating False Alarm Probabilities (FAPs) associated with the chosen measure is tricky even for traditional Fourier analysis of Gaussian white noise data \citep{Percival1993} and even more difficult when irregular cadences or autocorrelated noise is present. Issues concerning FAP estimation have been extensively discussed in the context of Lomb-Scargle periodograms (LSPs). Early analytic FAPs for LSPs \citep{Scargle82, Horne86} were shown to be unreliable \citep[e.g.,][]{Koen90, Czerny98}, and numerous improvements were suggested. Two broad approaches to estimating statistically significant periodic signals from periodograms have emerged: one based on the `sigma' noise level of the periodogram (e.g., \citealt{Scargle82}) and another based on periodogram extreme values. One outcome of these astrostatistical analyses is that procedures ignoring most periodogram noise values might better measure true periodicities. The statistical field of extreme value theory (EVT) based on the Fisher-Tippett-Gnedenko Theorem provides a mathematical foundation for evaluating peak significance in astronomical LSPs \citep{Baluev08}; we review the underlying mathematics in Appendix \ref{sec:evt}. 

EVT has been widely applied to problems in geology, finance, and engineering; for example, EVT helps evaluate whether a storm exceeds a `100-year hurricane' or whether a sudden stock market change is a fluctuation or a `crash'. The mathematics and many applications of EVT are presented in texts like \citet{Coles01} and \citet{Castillo04}.  The application of EVT to Lomb-Scargle periodogram FAPs has been discussed in astrostatistical studies \citep{Baluev08, Suveges14, Suveges15, Sulis17, Vio19, Delisle20, Koen21, Giertych22} and the review by \citet{VanderPlas18}. EVT is also gaining increased attention for other astronomical applications, including solar, stellar, galaxy, and cosmological studies \citep[e.g.][]{AsensioRamos07, Pratt17, Waizmann12, Davis11}.  We extend the application of EVT-based detection to periodograms specializing in planetary transit detection guided by the approach of \citet{Suveges14} as described in Appendix \ref{sec:evt}. 

\subsection{Scope of This Study}

The present study aims to investigate statistical issues related to the behaviors of BLS and TCF periodograms and their sensitivities to small planetary transit signals under different noise conditions.  Much of our effort is based on the analysis of simulated light curves described in \S\ref{sec:simulations}, where we discuss two metrics to evaluate the significance of periodogram peaks. This reveals previous properties of the two periodograms that extend the work of \citet{Ofir14} and show their dependencies on light curve properties (\S\ref{sec:simresults}-\ref{sec:results_LCprop}). We then illustrate the two periodograms on real TESS light curves with known small planets (\S\ref{sec:TESS}). After a discussion of the findings (\S\ref{sec:discussion}), the study ends with advice to transit survey scientists on the best approaches for small planet detection (\S\ref{sec:conclusion}). 

Our analysis is not intended to be a comprehensive study of periodograms for transit study. The effort here is limited to comparing the performance of two periodograms, BLS and TCF, applied to light curves with continuous evenly-spaced cadences. Our simulations described in \S\ref{sec:simulations} make specific assumptions about the transit shape and exclude some astrophysical effects. Both are low-dimensional parametric procedures with fixed functional forms at a chosen trial period: a rectangular box for BLS and a double-spike pattern for TCF.  We do not treat cases with heteroscedastic light curves where different data points have different weights, nor cases where the noise is non-Gaussian.

We do not consider other periodograms based on sinusoidal variations, such as the Schuster periodogram of classical Fourier analysis or its Lomb-Scargle extension to irregular observing cadences \citep{Scargle82}, nor do we consider nonparametric periodograms such as phase dispersion minimization \citep{Stellingwerf78} and minimum string length \citep{Dworetsky83}. The Transit Least Squares (TLS) algorithm \citep{Hippke19,Heller22}, an important variant of BLS  with ingress and egress transit shapes arising from astrophysical modeling, is not analyzed. Period search procedures with different statistical approaches are not treated, including \citet{Waldmann12} and \citet{Zucker15}.  We also ignore evaluations of periodicity significance in the time domain, such as the classical Wald test \citep{Pont06}. We assume datasets with evenly spaced time series (although missing data may exist). Thus, we do not consider highly irregular light curves typically emerging from ground-based telescopic surveys. Our calculations use the original BLS algorithm of \citet{Kovacs02}, and we only briefly mention faster algorithms in Appendix~\ref{sec:execution_time}. Comparison of period finding algorithms, although with different focuses, has been performed in previous studies (e.g., \citealt{Graham13} and references therein).

One method not examined here may have high sensitivity to small planets. \citet{Gregory92} formulate a likelihood for Bayesian period search assuming an arbitrarily shaped transit. A sensitive likelihood-based periodogram might emerge if one inserts a strong prior for a periodic box or double spike shape with a small duty cycle.

\section{Simulated Light Curves}
\label{sec:sim}

\subsection{Construction and Analysis of Simulated Light Curves}
\label{sec:simulations}

Simulations provide an excellent way to assess and compare periodogram peak significance due to full control over noise conditions. Our simulations are designed to roughly resemble single-sector observations from the TESS satellite prime mission survey. The light curves have a uniform 0.5 hr observing cadence. No gaps in observations are included.  The noise behaviors are stationary--no changes during the entire time span of the light curve are modeled.  In analogous to observed light curves, this implies that nonstationarity has previously been removed with a detrending procedure: moving average filter, spline or Gaussian Processing regression, wavelet transform, or similar operation. 

Two noise models are constructed.  The first assumes Gaussian white noise with mean 0 and standard deviation $\sigma_0 = 1 \times 10^{-4}$ or 100 parts-per-million (ppm).  This is characteristic of TESS photon noise levels for bright stars with $T \simeq 7-8$ where super-Earths might be detected around solar mass stars, and is similar to the simulation noise level assumed by \citet{Hippke19}.  

The second noise model assumes an autoregressive moving average (ARMA) process with order ARMA(3,3) and coefficients sufficiently high to give statistically significant autocorrelation up to $\simeq 5$ hours. Specifically, we assume the flux value $X_t$ at time $t$ is 
\begin{equation} \label{eqn:arma}
    X_{t} = \sum_{i = 1}^{3} \phi_i X_{t-i} + \sum_{i = 1}^{3} \theta_i \epsilon_{t-i} + \epsilon_{t}
\end{equation}
where $\epsilon_{t} = N(0,\sigma^2)$ is a white noise process with $\sigma = 1 \times 10^{-4}$ and the ARMA coefficients are set to $\phi = (0.2, 0.3, 0.2)$ and $\theta = (0.2, 0.2, 0.3)$\footnote{
Note that the ARMA model here is actually a superposition of autoregressive (AR) behavior, moving average (MA) behavior, and Gaussian white noise due to the $\epsilon$ scatter term. $\epsilon$ refers to the uncorrelated scatter in the time series arising from some combination of photon noise, instrumental problems, and intrinsic stellar variability.  We do not associate it with any specific cause. For background on ARMA modeling, see textbooks like \citet{Box15} and \citet{Chatfield19}. For notational ease in this paper, we call the composite model of equation (\ref{eqn:arma}) "autoregressive noise".  The autoregressive noise light curves were constructed using funtion `arima.sim' in the R statistical software environment.
}. While these coefficients may be larger than realistically present in many detrended TESS light curves, the simulations are designed to reveal clear differences between periodogram performance for white and correlated noise.  For reference, ARIMA models have been applied to large samples of Kepler and TESS light curves by  \citet{Caceres19b} and \citet{Melton23a}, respectively, based on the methodology described by \citet{Caceres19a} and \citet{Melton23b}.

Simulations with a hypothetical transiting planet contain periodic transits with varying characteristics such as the orbital period, transit depth, duration, noise type, and total transits (\S\ref{sec:results_LCprop}). The shape of the transits is modeled as a trapezoid with 30 min ingress and egress; limb darkening, impact factor, and other possible effects are not included. We simulate only one planet; systems with multiple transiting planets producing multiple periodic light curves are not examined. 

Before calculating the BLS periodogram when autocorrelated noise is simulated (but not when pure Gaussian white noise is simulated), the simulated light curves are subject to Gaussian Processes regression, which removes trends but may leave short-memory autocorrelation.  We use the software implementation $gausspr$ in CRAN package $kernlab$ \citep{Karatzoglou23} within the R statistical software environment \citep{RCoreTeam22} based on the methodology described by \citet{Karatzoglou04} and \citet{Williams98}. We use the squared exponential kernel (the radial basis kernel function) as the covariance function. The kernel width hyperparameter, $\sigma$, is set using a heuristic to set a reasonable value based on the data (controlled by the keyword \texttt{kpar = `automatic'})\footnote{Our internal experiments showed that changing $\sigma$ as 0.1, 1, and 10 yield only minor fluctuations in the significance analyses suggesting that our conclusions would not depend on the $\sigma$ value.}.

Before calculating the TCF periodogram, the simulated light curves are subject to ARIMA modeling that effectively removes both trends and short-memory autocorrelation, leaving residuals close to white noise. We use the software implementation $auto.arima$ in CRAN package $forecast$ that automatically calculates maximum likelihood fits for a range of (p,0,q) orders and selects the best model based on the Akaike Information Criterion that is penalized for model complexity \citep{Hyndman08, Hyndman23}.  We restrict model complexity to $p, q \leq 5$.

Periodograms are then calculated from the detrended light curves to reveal transit-shaped periodic behaviors. As outlined in \S\ref{sec:intro}, we will be examining the sensitivity of two periodograms $-$ BLS and TCF $-$ and two measures of significance in periodogram peaks: an SNR that takes into account trends and heteroscedasticity in periodogram noise; and an EVT-based probability of a standardized periodogram that ignores periodogram noise and considers only extreme values. Such a combination has been used since both metrics have their advantages and disadvantages: FAPs affected by aliasing--a common feature in astronomical periodograms (\citealt{Baulev13}; \citealt{Baluev08}; \citealt{Suveges14}); SNR affected by complex alias structures, non-Gaussianity \citep{Caceres19a}. These significance measures are described in \S\ref{sec:per_snr} and \S\ref{sec:per_EVT}, respectively.

\subsection{Periodogram Peak Metric Based on Signal-to-Noise Ratios}
\label{sec:per_snr}

It is not improbable that a periodicity from a small planet produces a peak in a periodogram that can not be unambiguously interpreted as a transiting planet due to the noise characteristics of the periodogram in surrounding frequencies. If the periodogram power values and noise variance are not heteroscedastic, this problem can not be effectively captured in bootstrap procedures as used in EVT-based analysis (\S\ref{sec:per_EVT}). However, this situation is treated by a local SNR measure providing the noise is estimated using frequencies close to the peak of interest. We emphasize that the noise considered here is in the frequency-domain periodogram, not the noise in the original time-domain light curve as considered in many other studies \citep[e.g.][]{Kovacs02, Pont06, Fressin13, Dressing15}.

To account for the periodogram trends (typically a rise in the mean level of the periodogram with the increasing period as noted by \citealt{Ofir14}), the periodogram is detrended using a smoother designed to be robust against non-Gaussianity and outliers. We use the median (50\% quantile) curve of a quadratic smoothing B-splines with roughness penalty parameter $\lambda=1$ using the method developed by \citet{Ng96} and \citet{Ng07}. Twenty equally-spaced spline knots are used. Code implementation is provided by CRAN package $cobs$ \citep{Ng22}.

We then define the SNR of the periodogram peak as
\begin{equation}
    \textrm{SNR} = \textrm{Power}_{peak} / \textrm{MAD}_{peak}
\label{eqn:SNR}
\end{equation}
where $\textrm{Power}_{peak}$ denotes the peak periodogram power, $\textrm{MAD}_{peak}$ is the median absolute deviation of nearby periodogram powers measured in a window of frequencies around the peak under study after the periodogram has been detrended (\S\ref{sec:simulations}).  

The MAD is a robust measure of local scatter that, unlike the usual root-mean-square $\sigma$ value, is insensitive to strong non-Gaussianity of periodogram noise values. The MAD is used in this context, for example, by \citet{Vanderburg16} in a search for K2 transiting planets. We note that there are no theorems associated with our definition of SNR, and its statistical distribution is unknown.

There is little guidance on defining a `nearby' region of the periodogram to measure the noise; we select a window of 3000 periods symmetrically centered around the peak. This choice is generally sufficiently large to give a good estimate of the MAD but sufficiently narrow to avoid heteroscedasticity (changes of noise amplitude with the period) in the periodogram.

\subsection{Periodogram Peak Metric Based on EVT False Alarm Probabilities}
\label{sec:per_EVT}

Mathematical background of extreme value theory (EVT) and its application for periodogram peak significance is given in Appendix~\ref{sec:evt}. Our approach of applying EVT (\S\ref{sec:evt}) to periodogram peak evaluation closely follows \citet{Suveges14}; see also \citet{Suveges12} and \citet{Suveges15}. Here a non-parametric bootstrap procedure is combined with EVT to estimate FAPs for the peak in a periodogram. The first step is generating $R$ bootstrap samples from the original time series. The next step is calculating the periodogram for each of the $R$ bootstrapped series and selecting the maximum of each periodogram, where each periodogram is computed on $K \times L$ frequencies selected from the entire frequency grid. This yields a sample of $R$ maximum values to which the GEV distribution is fit to obtain the FAP of the peak in the original periodogram. These $K \times L$ frequencies are selected by randomly selecting $L$ non-overlapping frequency intervals, each with $K$ consecutive frequencies. Such a selection ensures that long-range dependencies are accounted for due to $L$, while short-range dependencies (spectral leakage) are accounted for due to $K$.

We have conducted tests to ensure that the sensitivity to faint planets evaluated using our EVT procedures is largely independent of reasonable choices of $R$ ($100-500$), $K$ ($1-5$), and $L$ ($100-500$), similar to the stability experiments performed by \citet{Suveges14}. These tests were made for simulated light curves with Gaussian white noise and autocorrelated noise. One might also predict that unnecessarily higher oversampling might introduce spurious small-scale structures in the periodogram. This effect might be present for $K \simeq 10-20$ but is not seen in our simulations for $K \leq 5$.

A GEV distribution is then fit by maximum likelihood to the sample of $R$ maxima to obtain the FAP of the peak in the original periodogram. The quality of the GEV fit is checked with the Anderson-Darling test \citep{Stephens1974}, requiring $p$-value $> 0.01$. A similar ``shortcut" described in \citet{Koen15} was used: the three estimated GEV parameters were treated as known while calculating the $p$-value for these tests. This goodness-of-fit test is needed because the GEV is only a limiting distribution. If the model is deemed valid, it can be used for estimating the FAP of the observed peak in the periodogram. Here 
\begin{equation}
    \textrm{FAP} = 1 - \widehat{G}(x)
\end{equation}
where $\widehat{G}$ is the fitted GEV distribution, and $x$ is the peak power of the periodogram calculated on the original time series. Despite the common practice of oversampling periodograms, which introduces dependency, approaches based on EVT have shown plausible results, provided one has verified the GEV fit quality.

To facilitate the comparison of periodograms using FAPs, we apply the EVT procedure described above on ``standardized" periodograms, i.e., with its trend removed and normalized by the local scatter, that converts the periodogram powers to similar scales. The detrending is performed with the same procedure used for SNR calculation described in \S\ref{sec:per_snr}. The local scatter of the detrended periodogram is then estimated as a running MAD using ten windows in the entire frequency grid. The periodogram powers at the edge of the frequency grid are handled by taking their MAD. The local scatter implementation is taken from the \texttt{runmad} function from the \texttt{caTools} CRAN package version 1.17.1. We advocate the use of a ``local" scatter measure rather than a ``global" scatter measure used previously by, e.g., \citet{Hippke19}, to account for the heteroscedastic noise structure observed in the BLS and TCF periodograms \citep[][]{Ofir14, Caceres19a}. Standardization is also performed on the partial periodograms before extracting their maxima for GEV fitting.

As discussed in \S\ref{sec:evt}, it is important to note that this approach is not directly applicable to correlated time series. However, here the light curves have been detrended prior to periodogram applications when correlated noise is simulated and thus are close to uncorrelated. We remind the reader that detrending is needed twice: once on the light curves and second on the periodograms; however, both differ.

\section{Detailed Examination of a Single Trial of Transiting Planets}
\label{sec:simresults}

\subsection{Sensitivity comparison using FAP and SNR metrics}
\begin{figure*}[t]
\centering
\includegraphics[keepaspectratio,width=\linewidth]{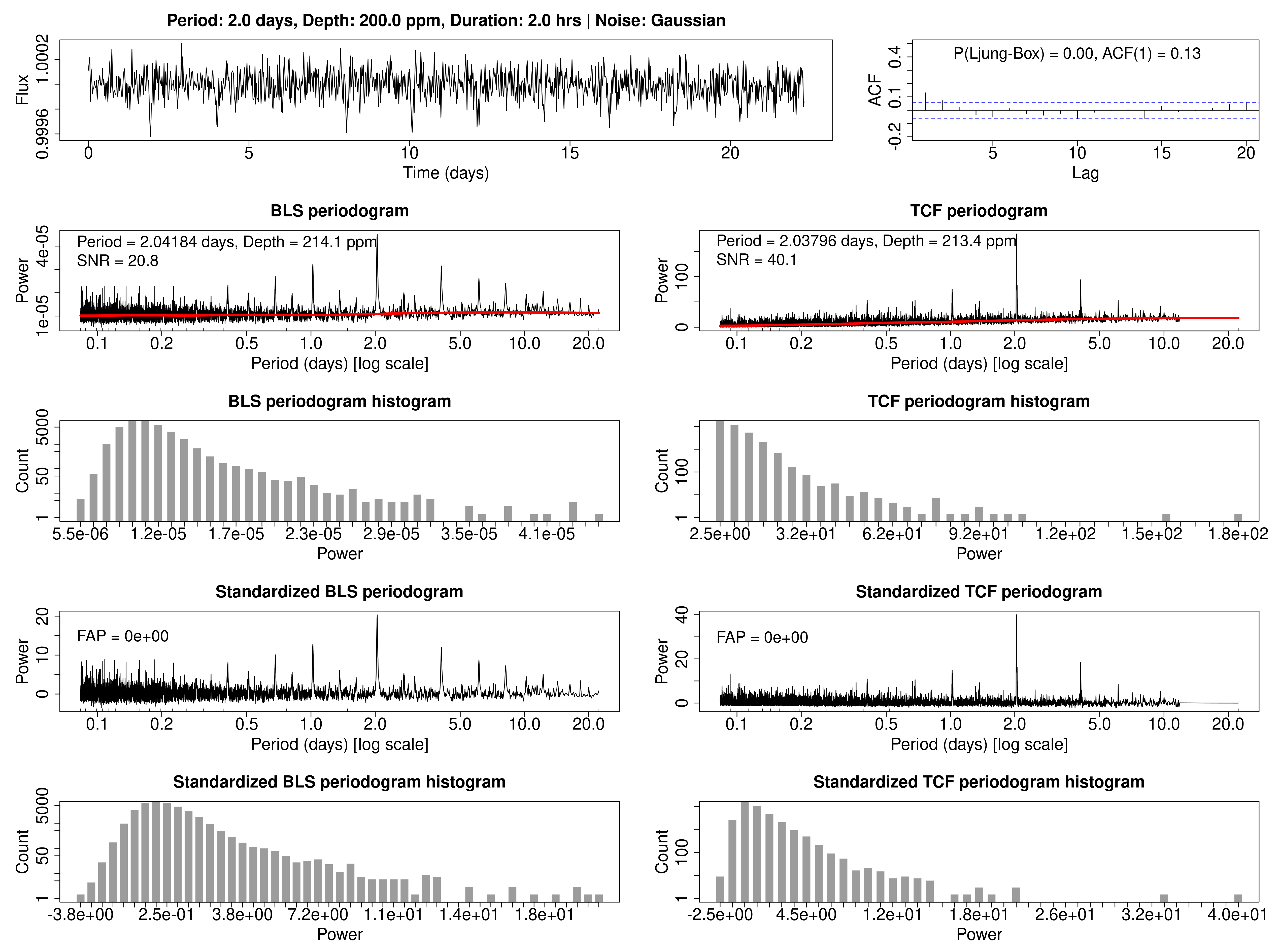}
\caption{Periodogram analysis of a simulated light curve with Gaussian white noise together with an injected planetary transit with depth 200~ppm. The light curve cadence and duration are similar to a single-sector TESS FFI observation. The injected planet has an orbital period of two days and a transit duration of two hours. The top row shows the simulated light curve (left) and its autocorrelation function (right). The second row shows the BLS (left) and TCF (right) periodograms. The red curve gives the $cobs$ median fit to the periodogram powers. A rug plot (small vertical marks on the x-axis) shows the location of the knots used for the median fit. Annotations give the SNR, period, and depth obtained from the periodogram. The third row gives histograms of the BLS (left) and TCF (right) power values. The fourth and fifth rows show the periodograms and the corresponding histograms after standardization. The annotation in the fourth row gives the FAP value of the peak in the standardized periodogram}
\label{fig:simGauss1}
\end{figure*}

\begin{figure*}[t]
\centering
\includegraphics[keepaspectratio,width=\linewidth]{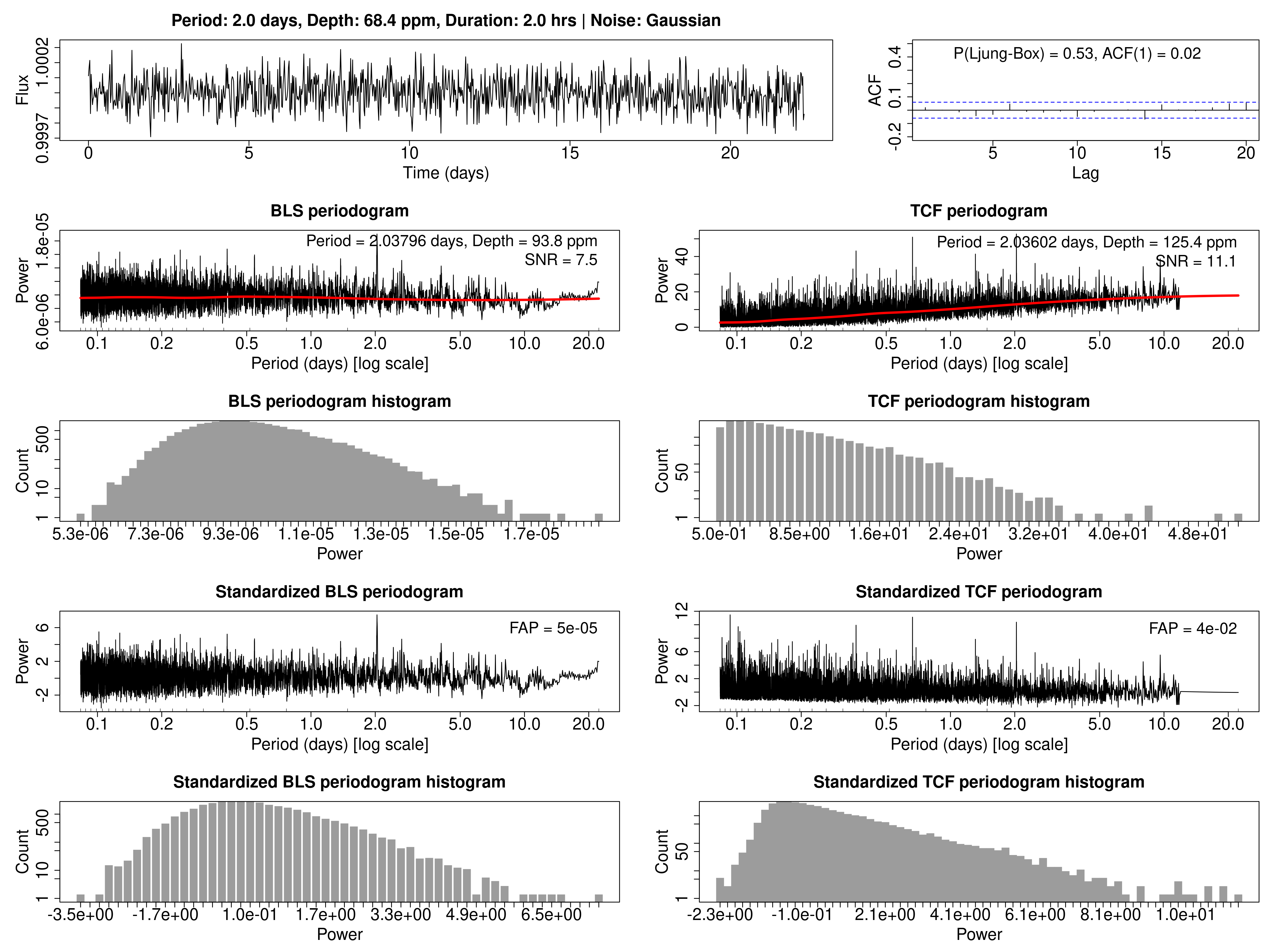}
\caption{Periodogram analysis for Gaussian  white noise and a smaller planet with injected transit depth 68~ppm. See Figure~\ref{fig:simGauss1} for a description of the panels.} \label{fig:simGauss2}
\end{figure*}

\begin{figure*}[t]
\centering
\includegraphics[keepaspectratio,width=\linewidth]{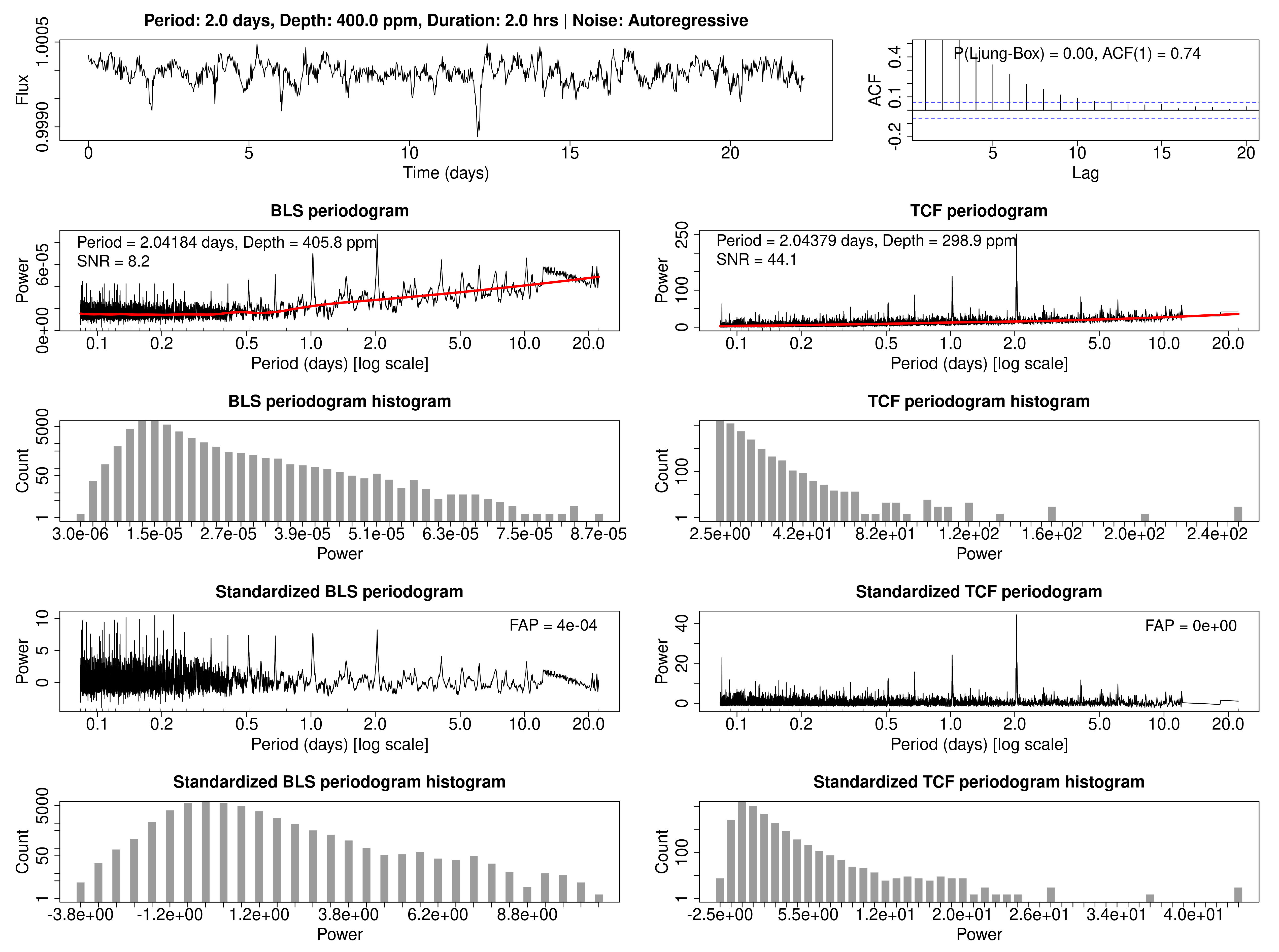}
\caption{Periodogram analysis with autocorrelated noise. Here the injected transit depth is 400~ppm. See Figure~\ref{fig:simGauss1} for a description of the panels.} \label{fig:simARMA1}
\end{figure*}

\begin{figure*}[t]
\centering
\includegraphics[keepaspectratio,width=\linewidth]{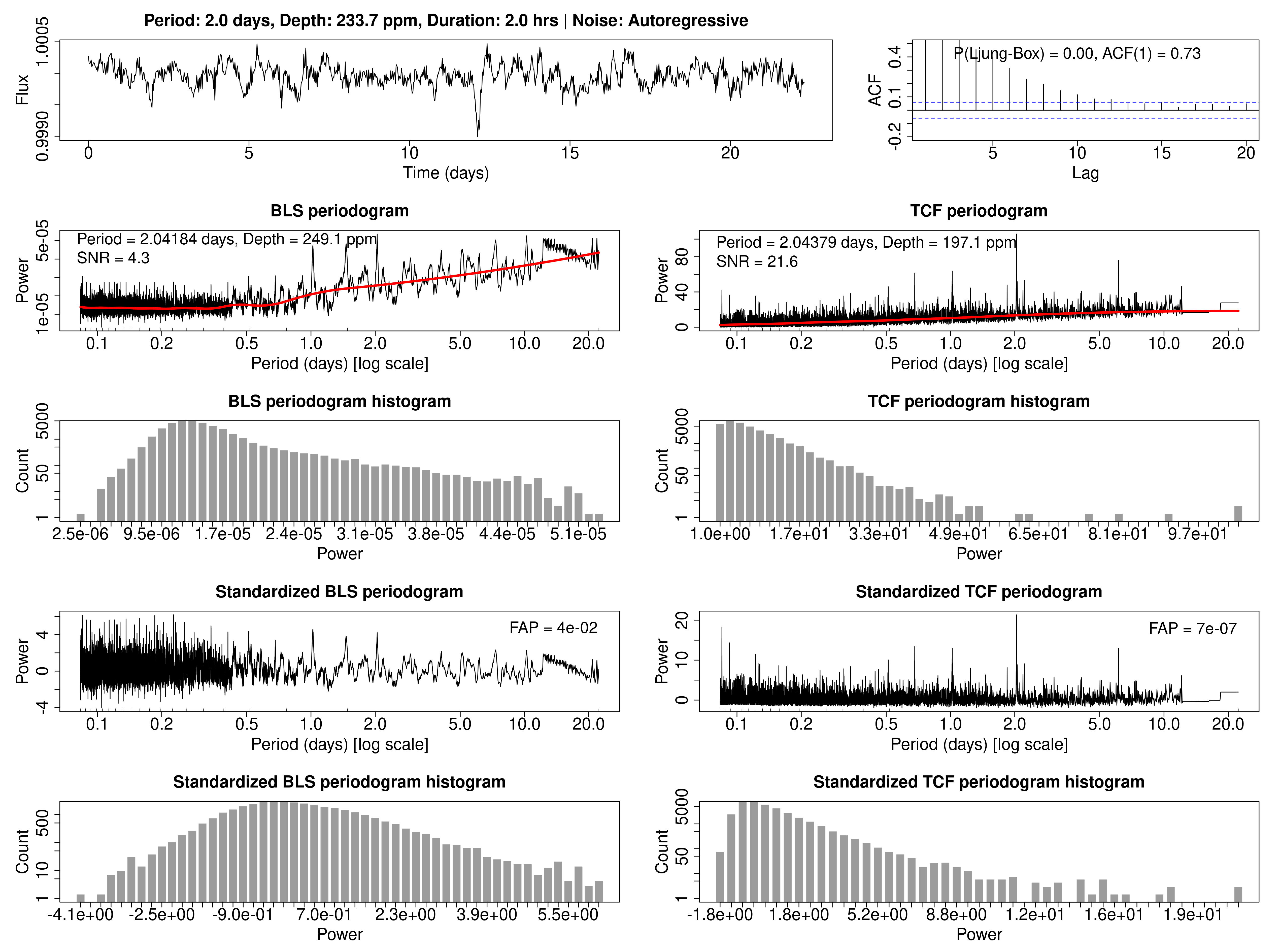}
\caption{ Periodogram analysis with autocorrelated noise and a smaller injected planet with 234~ppm. See Figure \ref{fig:simGauss1} for a description of the panels.} \label{fig:simARMA2}
\end{figure*}

We show the results on BLS and TCF periodograms with SNR and FAP peak evaluation metrics for simulated light curves with Gaussian and autocorrelated noise characteristics described above. For illustration here, we have injected transit signals for planets with two different sizes assuming an orbital period of two days and a two-hour transit duration. The results are shown in Figures~\ref{fig:simGauss1}-\ref{fig:simARMA2}.  



In the simulation with Gaussian white noise, the transits can be seen visually in the light curve for the larger planet (simulated depth = 200~ppm, Figure~\ref{fig:simGauss1}), but they are lost in the noise for the smaller planet (simulated depth = 68~ppm, Figure~\ref{fig:simGauss2}). The signal is recovered for the larger planet and only marginally recovered for the smaller planet by the BLS and TCF periodograms. However, the TCF gives a substantially stronger signal for Gaussian white noise: SNR = 40.1 $vs.$ 20.8 for the larger planet and 11.1 $vs.$ 7.5 for the smaller planet. Upward trends in noise values are seen in the TCF periodogram for longer periods (red $cobs$ curves) that are removed by standardization (fourth row). The BLS periodogram shows heteroscedasticity in noise as the period changes, as noted previously by \citet{Ofir14} and \citet{Caceres19a}. Similar to \citet{Ofir14}, we observe that the BLS scatter increases at longer periods (i.e., smaller frequencies) in the periodogram, which was found by observing an increasing trend in the scatter estimate described in \$\ref{sec:per_EVT}. Thus, an increasing trend in the power and the power scatter towards longer periods suggests that the distribution of powers at shorter and longer periods could differ. 

Although the TCF has greater sensitivity than BLS using the SNR metric due to a higher SNR, BLS is considerably more sensitive using the FAP metric. This is shown in the annotations in the fourth rows of Figures~\ref{fig:simGauss1}-\ref{fig:simGauss2} where $\textrm{FAP} \sim 5 \times 10^{-5}$ for BLS and $\sim4 \times 10^{-2}$ for TCF with the smaller simulated planet. For the larger planet, secondary peaks are mostly aliases of the injected 2.0~day period, while random noise peaks are the principal source of secondary peaks for the smaller planets.

Figures~\ref{fig:simARMA1}-\ref{fig:simARMA2} show the periodogram analysis using autocorrelated noise. It is difficult to unambiguously see the 2-hour transits visually in the light curve, even for a planet with depth = 400~ppm, twice the depth needed for a similar SNR for TCF when only Gaussian white noise was present. With autocorrelated noise, the ARIMA fitting and the TCF periodogram give SNR = 44.1, while the BLS periodogram is severely degraded with SNR = 8.2. Here a strong trend in BLS periodogram power levels for periods without true signal is present. Standardization of the BLS periodogram removes this strong trend (fourth row of Figure~\ref{fig:simARMA1}). However, the FAP value of $4 \times 10^{-4}$ is now worse than for the TCF periodogram. The TCF periodogram possesses only mild trends, unlike BLS. 

The situation for the smaller planet in the presence of autocorrelated noise is similar to the larger planet (Figure~\ref{fig:simARMA2}). The TCF periodogram captures the injected periodic transit without difficulty ($\textrm{SNR} = 21.6$), while it is hardly detected in the BLS periodogram ($\textrm{SNR} = 4.3$). The FAP metric gives a significant detection for TCF ($\textrm{FAP} = 7 \times 10^{-7}$) but an insignificant detection for BLS ($\textrm{FAP} = 4 \times 10^{-2}$). BLS shows strong trends and heteroscedasticity in the periodogram, while these problems are milder for TCF. Altogether, the ARIMA fitting and TCF periodogram are much better behaved than the BLS periodogram in the presence of autocorrelated noise. Thus, FAP and SNR present dissimilar conclusions regarding peak significance for Gaussian white noise but similar conclusions for autocorrelated noise.

We thus see a big difference in periodogram noise characteristics in response to light curves with Gaussian white noise $vs.$ autocorrelated noise. The BLS and TCF periodograms share similar properties $-$ mild trends and heteroscedasticity $-$ for Gaussian white noise. The TCF periodograms have a similar structure even for autocorrelated noise, as the autoregression is effectively removed by the ARIMA modeling that precedes TCF. However, Gaussian Processes (or similar nonparametric local) regression applied before BLS leaves significant autocorrelation if it was present in the original light curve. The BLS periodogram in Figures~\ref{fig:simARMA1}-\ref{fig:simARMA2} thereby exhibit undesirable strong behaviors not present in BLS periodograms in Figures~\ref{fig:simGauss1}-\ref{fig:simGauss2} with Gaussian white noise. Consequently, BLS has less ability to detect small planets if autocorrelation in the light curve still persists after detrending.

We thus find, for this single simulation, that TCF detects small planets more effectively $both$ for simulated light curves with Gaussian white noise (Figures \ref{fig:simGauss1}-\ref{fig:simGauss2}) and with autocorrelated noise (Figures \ref{fig:simARMA1}-\ref{fig:simARMA2}).

The FAP and SNR values shown in the plots correspond to a single realization of noise used to create the light curves. In practice, we have observed that the FAPs change when different noise realizations are used (see \S\ref{sec:resultsSetup} for more details).

\subsection{Supplemental periodogram analyses}

We can finally inquire into the accuracy of the transiting planet depth obtained from the BLS and TCF periodograms. For Gaussian noise, the depth estimates in both periodograms overestimate the simulated depths. The situation differs for autocorrelated noise, where the TCF underestimates the true depth. Inaccuracies in TCF depth estimation may have several causes: (a) incorporation of some transit signal into the ARMA fit can reduce the estimated depth; (b) ``over-differencing" by ARIMA can produce an anti-correlation at lag = 1 and increase the estimated depth; and (c) inaccurate registration of the cadence with respect to the transit ingress and egress can reduce the double-spike signal and estimated depth. See Figure 7 in \citet{Melton23b} and also \citet{Melton23a} for more discussion.

These problems with TCF depth estimation were noted by \citet{Caceres19b} and \citet{Melton23a} in their Kepler and TESS applications, which required improvement during the vetting phase of analysis. Difficulties with transit depth estimation also arise in BLS transit fitting, as discussed by \citet{Kovacs02} and \citet{Ofir14}. Altogether, planet parameters derived from periodograms alone may be inaccurate in complicated ways. 

As expected from the mathematical discussion in \S\ref{sec:evt}, one can visually see in rows 4-5 of Figures~\ref{fig:simGauss1}-\ref{fig:simARMA2} that the FAP significance calculation based on EVT depends on the rightmost tail region of the periodogram power distributions. The histograms also provide information about the overall noise characteristics of the periodogram. Ideally, for a single true periodic signal, the rightmost bin in the histograms would be a single isolated count\footnote{Since we oversample the periodogram with $K=2$ in these simulations, often the principal peak has $\textrm{Count} = 2$ rather than 1 in the rightmost (or maximum) bin of the periodogram histograms.}. The presence of histogram values crowded near the maximum indicates that spurious peaks confusing the true signal are present, either from aliases or noise. The shape of the histograms for the bulk of noise values of the BLS and TCF periodograms differ considerably; both have strongly skewed, non-Gaussian distributions. However, this distribution pattern does not affect the FAP significance analyses. 

An effect worth pointing out in the above figures is that post-standardization, the local periodogram noise at shorter periods is enhanced, which arises due to the heteroscedastic noise pattern of the periodogram. A modified windowed approach could be used to mitigate this issue to some extent; however, we do not deal with it here since the FAP calculation does not consider the periodogram noise.

A few other remarks are as follows:
\begin{enumerate}[nosep]
    \item We have also verified the $P^{-1/3}$ and $P^{-1/2}$ dependence of BLS and TCF periodogram peak, where $P$ is the planet period, as described in \citet{Caceres19a}. To achieve this, we simulated three planets with the same transit depth and duration but varying the period and number of transits so that the total length of the light curve remains approximately constant\footnote{Note that the dependence mentioned in \citet{Caceres19a} was concerned with real observational scenarios. Thus, increasing the period should be accompanied by decreasing the number of transits in our simulation code to ensure that the total length of observation remains constant. This differs from the experiments performed in Figure~\ref{fig:periodAndDurationCompare}.}. The BLS and TCF periodogram peak powers scale as 1:0.48:0.21 and 1:0.41:0.14, which is approximately similar to $P^{-1/3}$ and $P^{-1/2}$, respectively.
    \item Another notable observation in the autoregressive case (Figures~\ref{fig:simARMA1}--\ref{fig:simARMA2}) is that the widths of the BLS periodogram peaks are larger than TCF.
    \item It is important to note that the comparison of BLS and TCF presented in this paper is, in fact, a comparison of (a) Gaussian Processes regression + BLS and (b) ARIMA + TCF instead of BLS and TCF themselves. One might argue that a more faithful comparison should consider (a) ARMA + BLS and (b) ARIMA + TCF so that the actual behaviors of periodograms are apparent. However, we have found ARMA + BLS ineffective in detecting small planets, likely because ARMA fits typically capture the transits along with the non-transit, stellar autocorrelation. ARIMA + TCF does not suffer from this issue since there are negligible transit points to fit an ARMA model on the differenced light curve.
\end{enumerate}

\section{Periodogram Sensitivities for a Range of Planet Properties}
\label{sec:results_LCprop}
\subsection{Calculations}\label{sec:resultsSetup}

We generate simulated light curves for a range of planet and light curve properties following procedures described in \S\ref{sec:simulations}. All light curves in this section have a 0.5~hr cadence. Noise characteristics are white Gaussian noise or autoregressive noise following equation (\ref{eqn:arma}). Injected planets have trapezoidal transits with 0.5~hr ingress and egress. Computation of the BLS periodogram is a modified version of the original Fortran'77 BLS routine that accounts for edge effects and uses binning for computational efficiency \citep{2016ascl.soft07008K}. The TCF implementation is the Fortran code at \citet{2022ascl.soft06002C} with minor modification\footnote{This code can be obtained at \url{https://github.com/Yash-10/Periodogram-Comparison-Optimize-Planet-Detection}}. The SNR metric is calculated using equation \ref{eqn:SNR}, and the FAP metric is calculated as described in \S\ref{sec:per_snr} with $K = 2$, $R = 300$, and $L = 300$. BLS and TCF periodograms are calculated using uniform frequency sampling\footnote{We use constant frequency intervals in our periodograms instead of frequency intervals dependent on frequency as used by \citet{Ofir14} and \citet{Caceres19a}. This choice wastes some computational resources but should have little effect on the accuracy of the periodograms.}.

To compare the performance of periodograms under different conditions, we define a threshold called ``minimum detectable depth'' (MDD) of a small planet based on the SNR and FAP metrics. No clear consensus has emerged in the research community on the best thresholds that balance sensitivity for small planet detection against False Alarm reports. SNR thresholds used for (often standardized) BLS periodograms include  $\textrm{SNR} > 6$ \citep{Kovacs02}, $\textrm{SNR} > 15$ \citep{Ofir14}, $\textrm{SNR} > 9$ \citep{Vanderburg_2016}, and $\textrm{SNR} > 5$ \citep{Shallue_2018}. FAP thresholds are widely used for planet detection using Lomb-Scargle or BLS periodograms with values typically ranging over $0.001 < \textrm{FAP} < 0.01$ \citep[e.g.,][]{Maxted_2011, Lund_2015}. The threshold $\textrm{FAP} = 0.003$ corresponding to the Gaussian $3\sigma$ criterion lies in this range.

Since our scientific goal here is to maximize sensitivity for small planets and not to minimize False Alarms, we use relatively low thresholds here\footnote{The non-parametric bootstrap + GEV procedure described in \S\ref{sec:per_EVT} yields a conservative FAP estimate (higher FAPs than expected), so it is naturally beneficial for optimizing small planet detection}. We define the MDD of a small simulated planet as the transit depth at which $\textrm{SNR} > 6$ or $\textrm{FAP} < 0.01$ depending on whether FAP or SNR is used. We calculate MDD values for chosen properties of the light curve or planet properties with trial planet injections of different depths, and the MDD is the lowest depth at which the planet depth is still significant, as quantified by the FAP or SNR of the periodogram peak. Our internal tests have shown that, for some cases, the FAPs of periodogram peaks change considerably across different noise realizations of the light curve but are stable in many other cases. The instability of FAPs across different realizations poses no issues for very high or very low FAPs but only for FAPs near the set threshold, 0.01. Subsequently, the MDD values are averaged across ten distinct noise realizations in all cases to get more reliable sensitivities. Planet properties cover the range of Kepler-discovered planets that might be detected in a single sector of TESS observations. 

We note that some of our simulations have the number of transits as a tunable parameter, as this most clearly reveals differences between periodogram performance. This differs from observational surveys, where the availability of time for observing a host star, rather than the number of transits, is known beforehand.

\subsection{Results}\label{sec:simResults}

\begin{figure*}[ht]
    \centering
      \includegraphics[keepaspectratio,width=\textwidth]{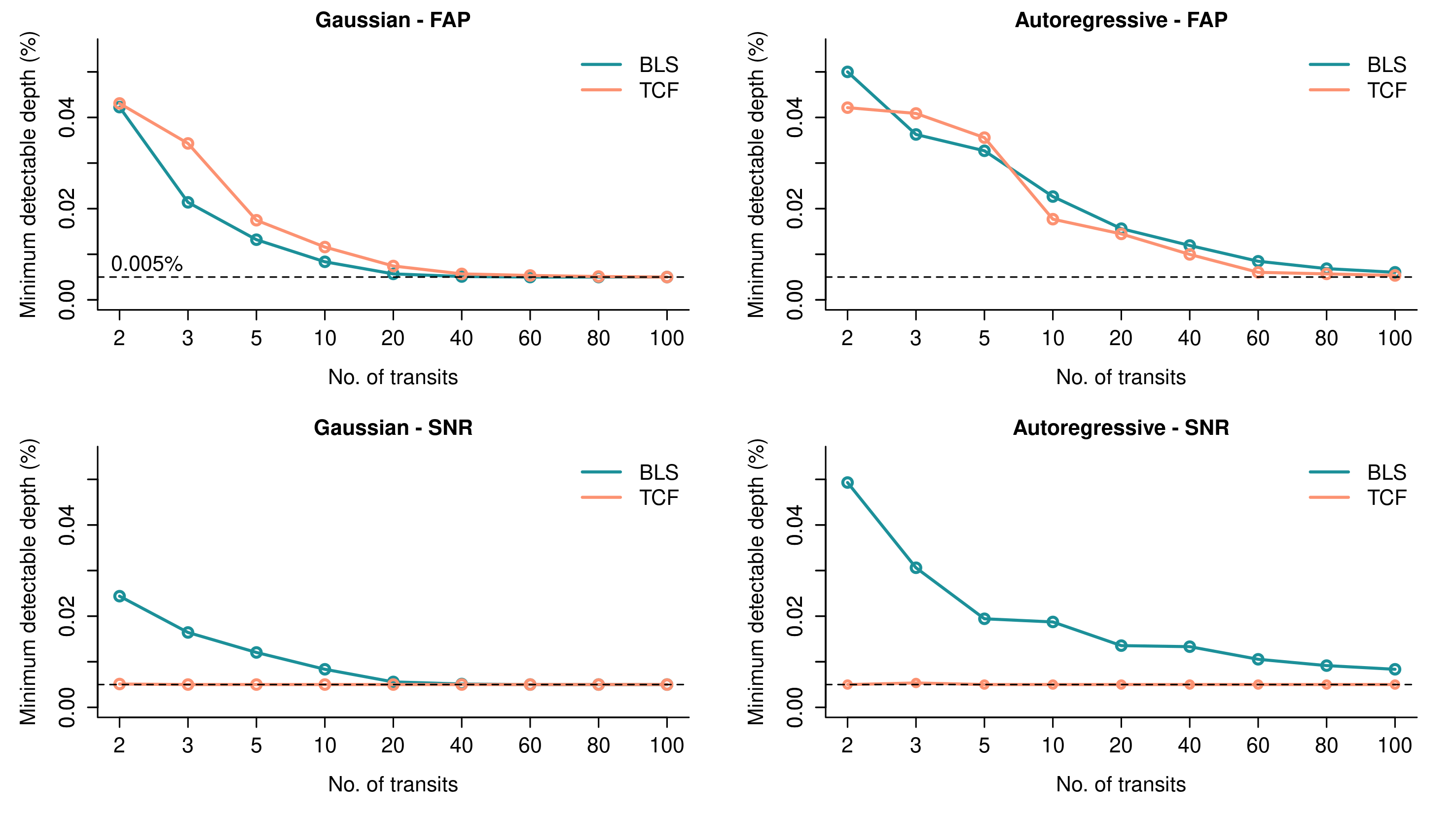}
    \caption{Minimum detectable depth (MDD) (in percent of the stellar brightness) as a function of the number of transits. The injected planet has a one-day period and a two-hour transit duration. Two metrics are shown: FAP based on extreme value theory on standardized periodograms (top row) and signal-to-noise ratio on detrended periodograms (bottom row). Light curves have Gaussian white noise (left column) and autocorrelated noise (right column). Each panel shows the MDDs for the BLS (blue curve) and TCF (orange curve) periodograms. The horizontal dashed line corresponds to MDD = 0.005\% to guide the eye.} 
    \label{fig:ntransitsBLSTCF}
\end{figure*}

\begin{figure*}
    \centering
      \includegraphics[keepaspectratio,width=\linewidth]{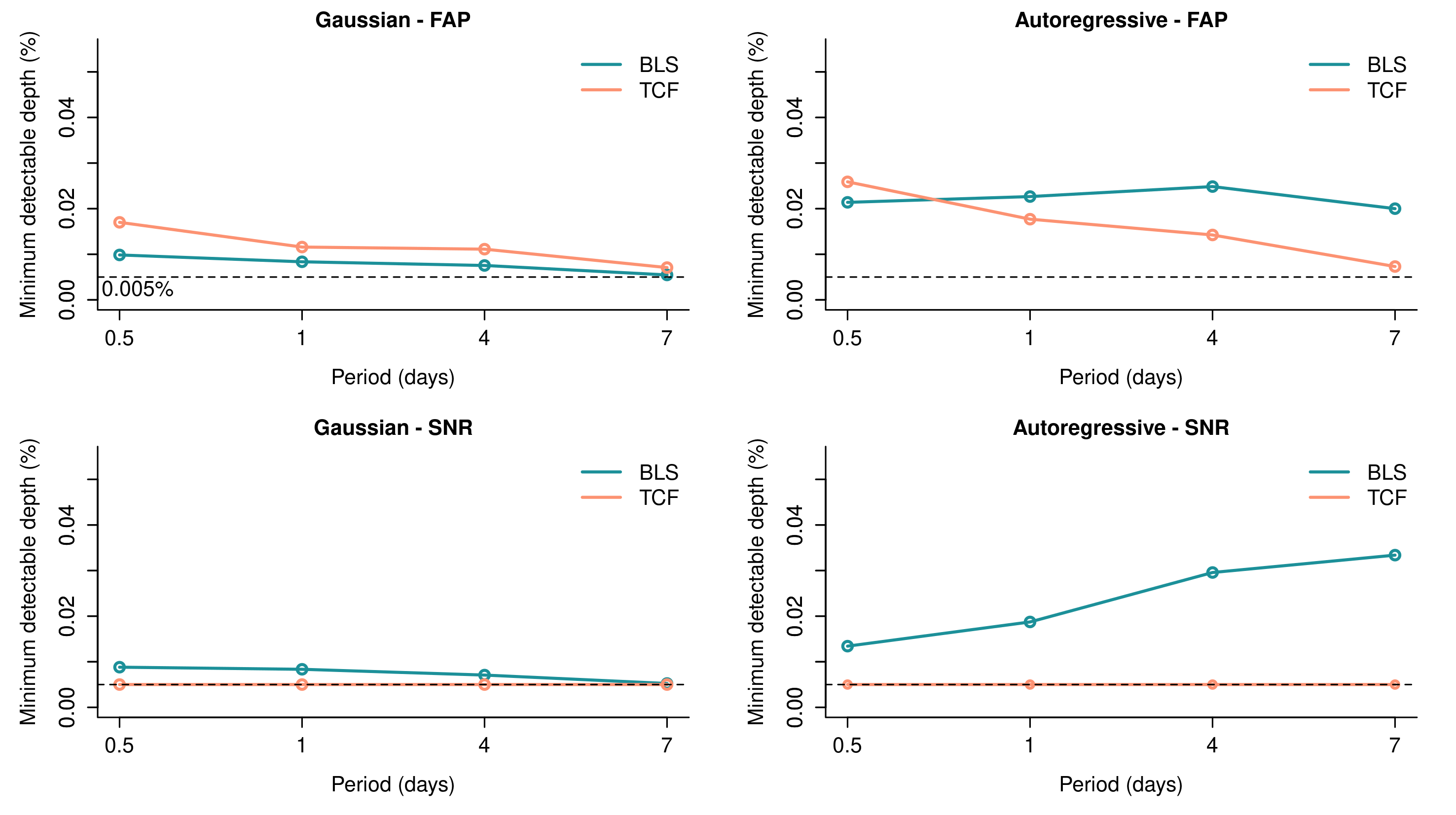}
      \includegraphics[keepaspectratio,width=\linewidth]{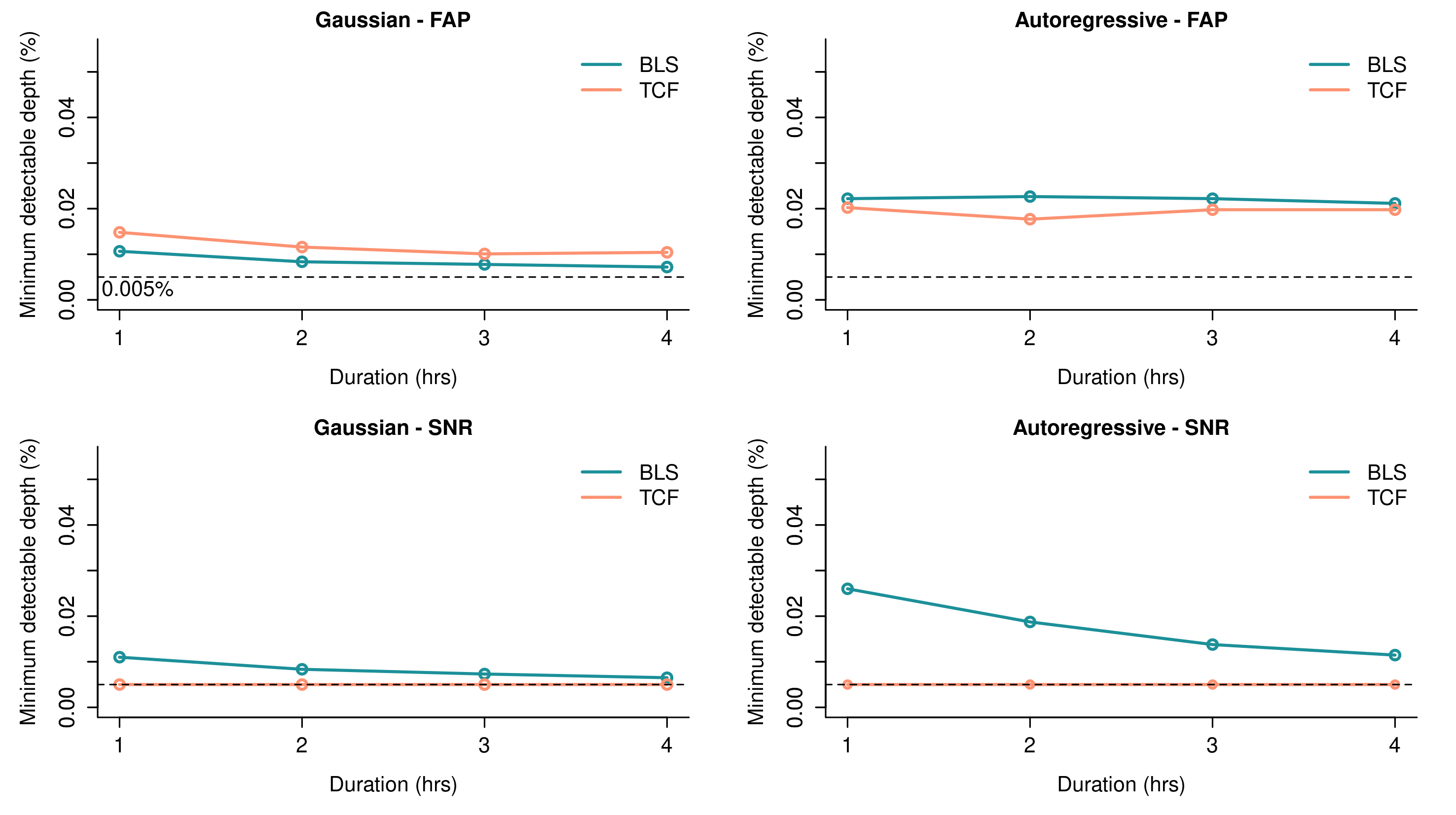}
    \caption{Minimum detectable depth (MDD) as a function of transit period (top rows) and duration (bottom rows). Simulations assume ten transits, transit duration = 2~hrs for the comparison using period, and period = 1~day for the comparison using transit duration. See Figure~\ref{fig:ntransitsBLSTCF} for panel details.} \label{fig:periodAndDurationCompare}
\end{figure*}

Figures~\ref{fig:ntransitsBLSTCF}-\ref{fig:periodAndDurationCompare} compare BLS and TCF as a function of the properties of the light curve (number of transits, white Gaussian $vs.$ autocorrelated noise), the properties of the injected planets (orbital period, transit duration) and the statistical metric for planet detection (SNR $vs.$ FAP).

The four panels of Figure~\ref{fig:ntransitsBLSTCF} give insight into a critical effect discussed in \S\ref{sec:simresults}: the sensitivity of the BLS and TCF periodograms is reversed depending on the nature of the light curve noise and chosen detection metric. For Gaussian white noise and the FAP metric (upper left panel), BLS is more sensitive than TCF (the blue curve lies below the orange curve). The periodogram sensitivities are similar for autocorrelated noise using the FAP metric, while TCF is considerably more sensitive when the SNR metric is used. Differences are most substantial when fewer transits are present; the choice of periodogram and detection metric becomes unimportant for short orbital periods embedded in long-duration light curves since the number of transits in such cases is sufficiently large.

As expected, sensitivity to small planets improves by extending light curves to include more transits; more points are available for building up the box-like transit signal for BLS and the double-spike signal for TCF. However, the improvement ceases after sufficient transits are observed, as both periodograms reach a fixed MDD. In these simulations with Gaussian white noise $\sigma = 100$~ppm, we set this limit to $\textrm{MDD} \simeq 50$~ppm (horizontal dashed lines). TCF's periodogram peak has an SNR more significant than the threshold, 6, even with two transits, as seen in Figure~\ref{fig:ntransitsBLSTCF}. A larger SNR threshold could increase TCF's MDD for the two-transit case. Overall, the FAP metric is less effective than the SNR metric for small planet detection when the number of transits is small.

Figure~\ref{fig:periodAndDurationCompare} compares the sensitivity of the BLS and TCF periodograms to the orbital period and transit duration. For most combinations of light curve noise characteristics and detection metrics, the MDD does not exhibit strong dependencies on these orbital properties. BLS's sensitivity shows some benefit from longer-duration transits; this effect is expected as BLS has more points to fit the box with its least squares algorithm. The TCF algorithm considers only the double spikes from ingress and egress and is thus not sensitive to transit duration (provided the period and number of transits are fixed).

The primary trend seen in Figure~\ref{fig:periodAndDurationCompare} is the deterioration in MDD for the BLS periodogram in the presence of autocorrelated noise (blue curve in the right panels, first and second rows). We suspect that this arises from the timescale of the autoregressive component we added to the light curve noise compared to the timescale of the orbital period. For $P = 0.5$~days, the variability structure covers a wide duty cycle in the folded light curve. However, for $P = 7$~days, the structure is confined to a narrow range of phases that mimic planetary transits. This points to the importance of effectively removing correlation in the light curve on timescales comparable to a transit duration. On the other hand, the ARIMA + TCF procedure removed the short-memory autocorrelation and ignored the relationship between period and transit duration. It thus has near-optimal MDD performance for the full range of periods for the SNR metric (orange curve in the right panel, second row). TCF benefits from longer-period transits using the FAP metric. In contrast, BLS only benefits when the noise is white Gaussian and instead deteriorates for autocorrelated noise (orange and blue curves in the two columns in the top row for TCF and BLS, respectively). Overall, the number of transits, not the period and duration (when the number of transits is fixed) is the dominant factor that affects the sensitivity of the BLS and TCF periodograms.

\section{Application to TESS light curves}
\label{sec:TESS}

To complement our comparison of the BLS and TCF periodograms in simulated lightcurves (\S\S\ref{sec:sim}-\ref{sec:results_LCprop}), we apply the procedures to four TESS FFI light curves drawn from the DTARPS-S survey \citep{Melton23b, Melton23a, Melton23c} that contain true known small exoplanets. Trends in these light curves have been removed using splines; however, we still preprocess the light curve using Gaussian Processes regression to maintain an analysis procedure similar to our simulations above\footnote{We observed negligible difference in the final results when Gaussian Processes regression is omitted here.}. We conduct the analysis without using the known period. 

The test performed here uses realistic rather than simplified light curves, no control over the noise level or characteristics, and gaps in observations from satellite operations. Since ARIMA requires uniformly-spaced time series, one can consider the observations evenly spaced with missing data points \citep{Feigelson18}.  The data gaps create spurious structures in periodograms, which may particularly affect GEV fits and associated FAPs \citep{Suveges14}. 

\begin{figure*}[ht]
\centering
\includegraphics[keepaspectratio,width=\linewidth]{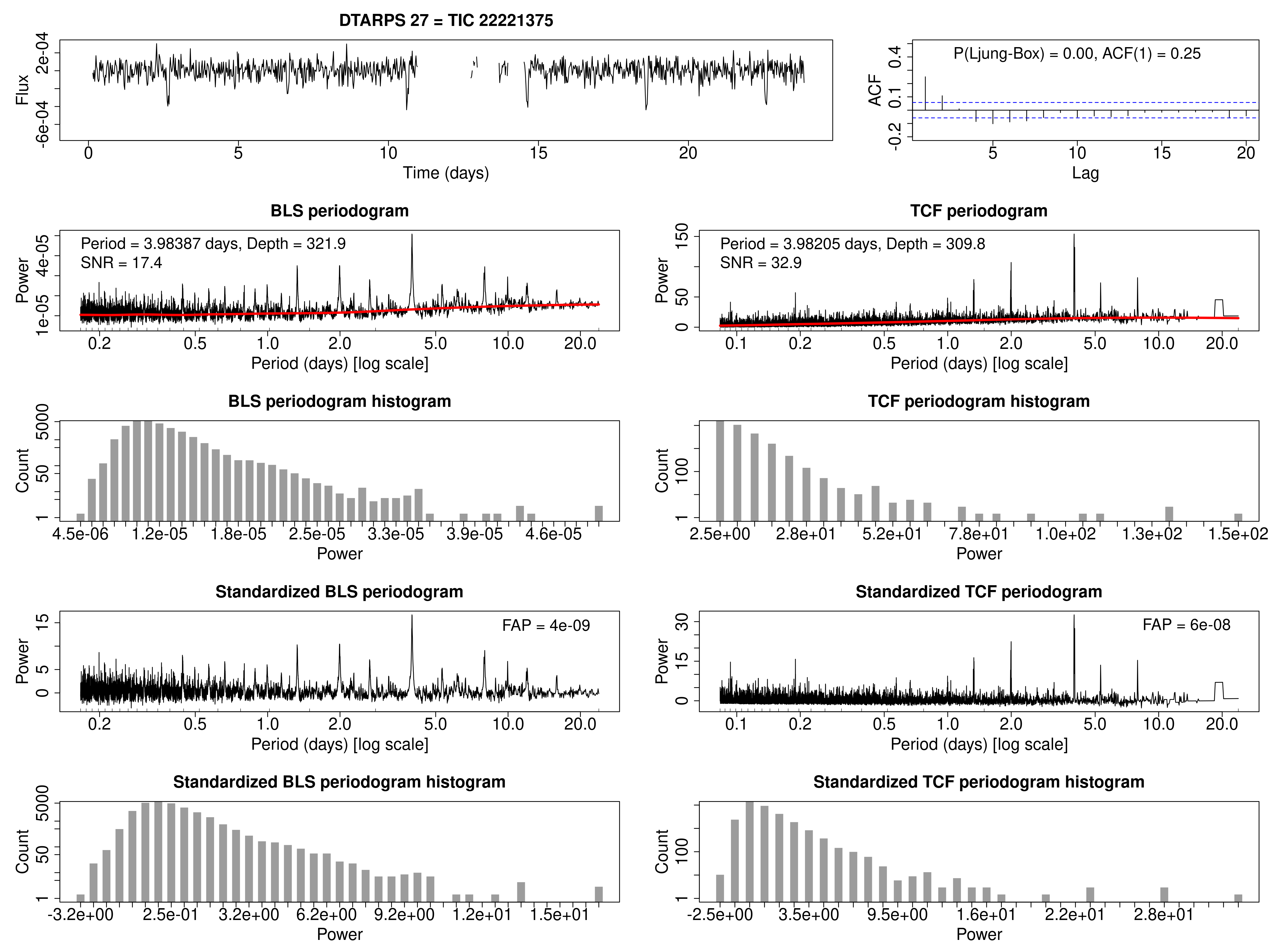}
\caption{Example 1 of a TESS Year 1 light curve for a star with smaller confirmed transiting planets. The panels are similar to those in Figure~\ref{fig:simGauss1} with the normalized light curve, autocorrelation function, BLS and TCF periodograms with median trend fit and their histograms, standardized periodograms, and histograms. Scalar results are provided in Table~\ref{tbl:4TESS}.} \label{fig:real1}
\end{figure*}

\begin{figure*}[ht]
\centering
\includegraphics[keepaspectratio,width=\linewidth]{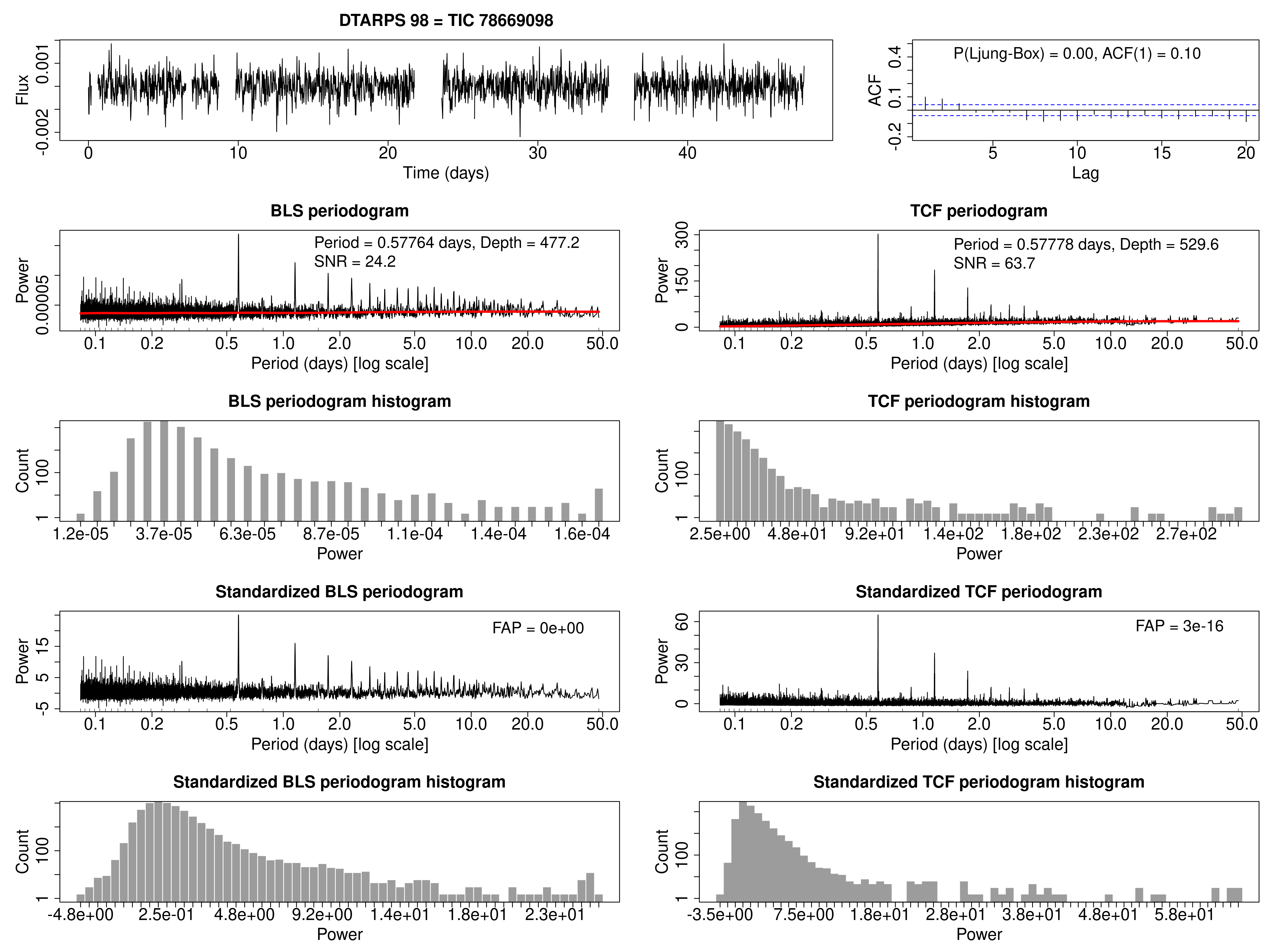}
\caption{Example 2 of a TESS Year 1 light curve and the corresponding periodograms and histograms. See Figure \ref{fig:real1} for a description of the panels.} \label{fig:real2}
\end{figure*}

\begin{figure*}[ht]
\centering
\includegraphics[keepaspectratio,width=\linewidth]{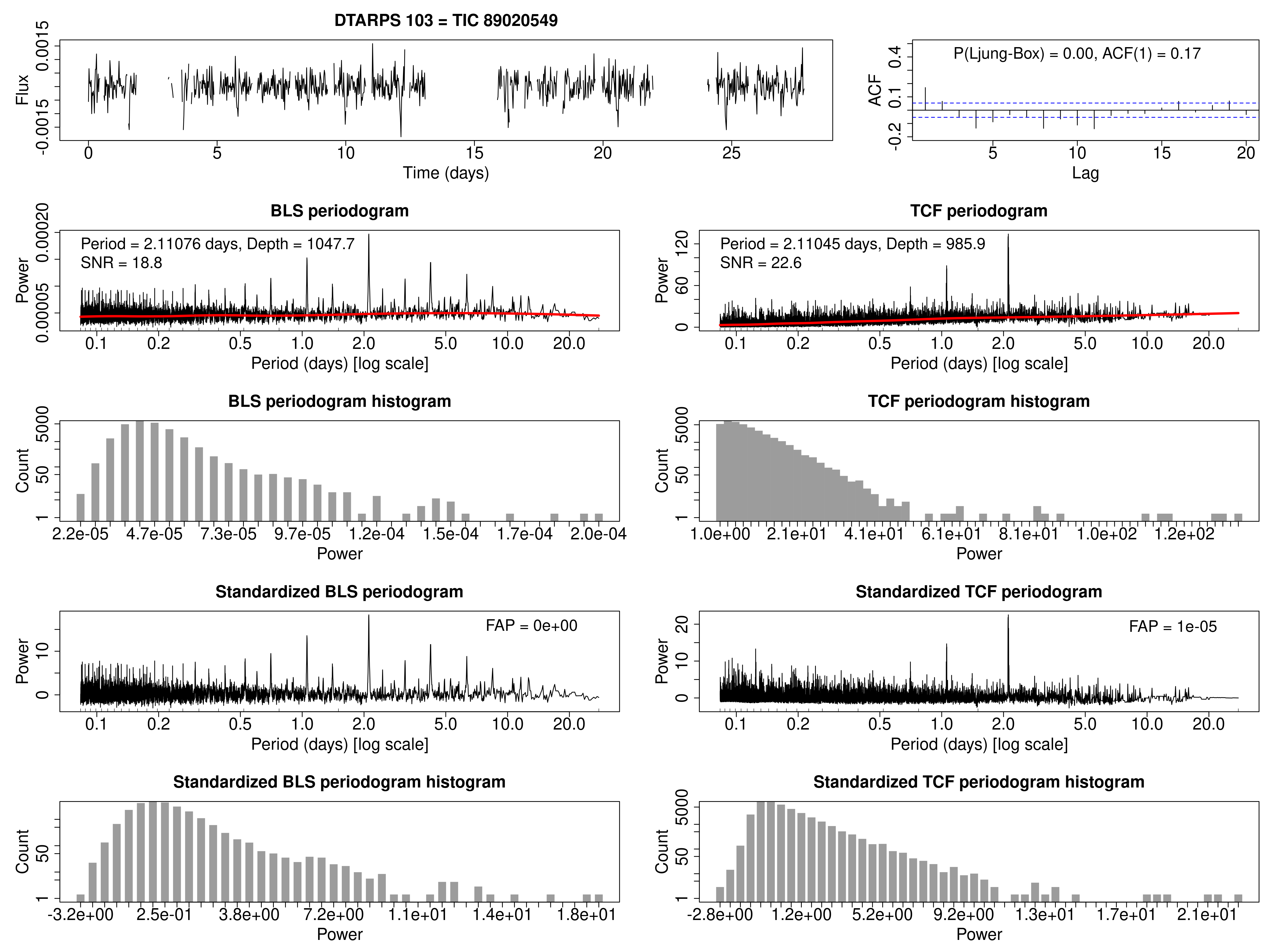}
\caption{Example 3 of a TESS Year 1 light curve and the corresponding periodograms and histograms. See Figure \ref{fig:real1} for a description of the panels.} \label{fig:real3}
\end{figure*}

\begin{figure*}[ht]
\centering
\includegraphics[keepaspectratio,width=\linewidth]{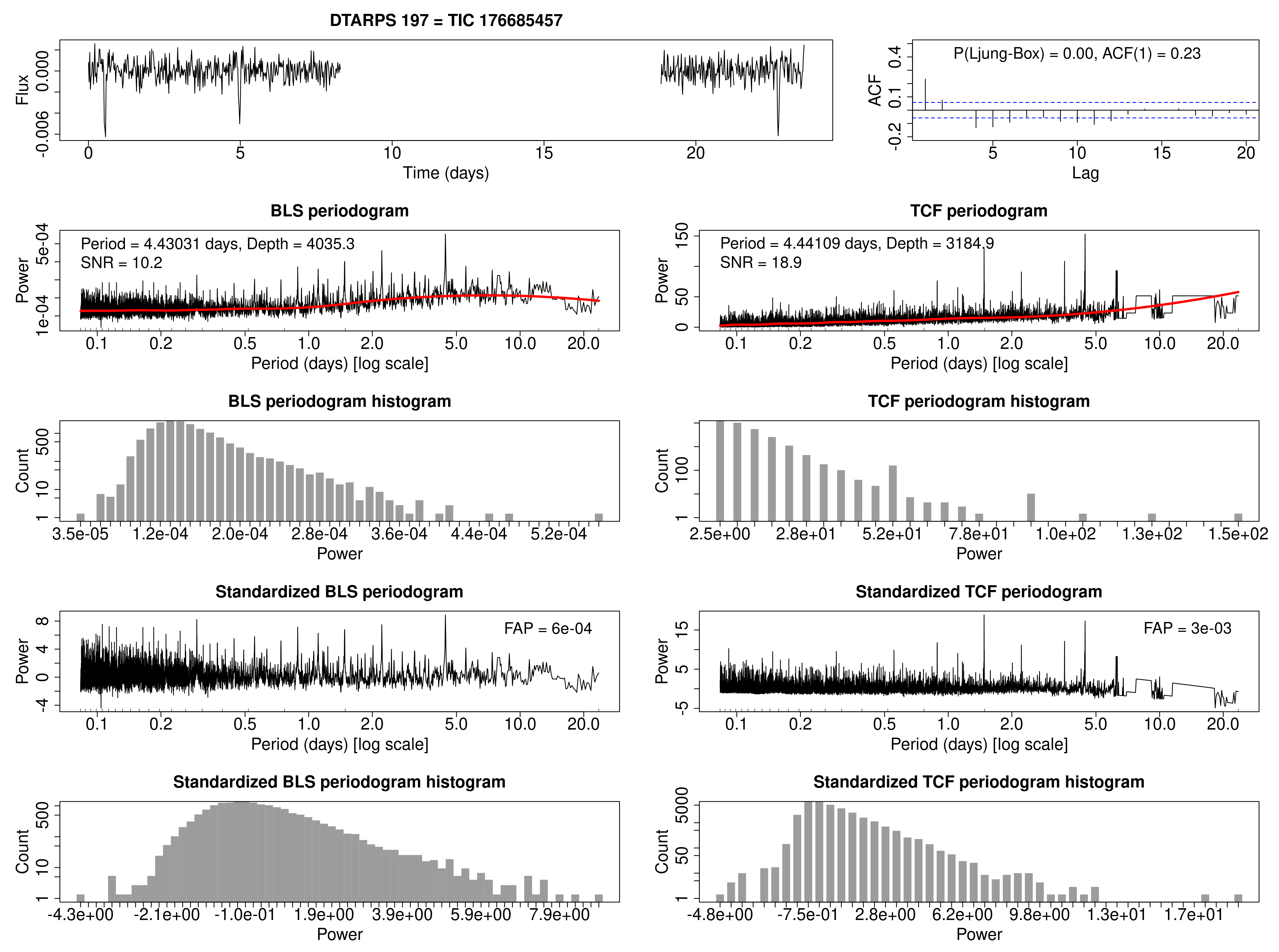}
\caption{Example 4 of a TESS Year 1 light curve and the corresponding periodograms and histograms. See Figure \ref{fig:real1} for a description of the panels.} \label{fig:real4}
\end{figure*}

Figures~\ref{fig:real1}-\ref{fig:real4} and Table~\ref{tbl:4TESS} show the results of the periodogram analysis on the four TESS light curves. One can see that both BLS and TCF periodograms obtained spectral peaks at the true orbital periods in all four cases. Other effects in Table~\ref{tbl:4TESS} are very similar to those found in the simulations. BLS and TCF estimated transit depths tend to underestimate true depths. All peak SNR values are much higher for TCF than BLS, while peak FAP values are significant for both BLS and TCF in all four cases.  

The autocorrelation functions in Figures~\ref{fig:real1}--\ref{fig:real4} show mild anticorrelations with lags up to 10 hours. Consequently, the periodograms show only mild trends and heteroscedasticity with the period. The expected aliases associated with the true period are seen in both periodograms. The TCF periodogram generally has a lower noise, giving it a higher peak SNR than BLS. The mild autocorrelation observed in these four cases also suggests that the expected comparisons should follow more closely to the Gaussian white noise simulations than the autoregressive noise simulations, which are observed here.

We thus find a complete validation of the simulation results in these real TESS FFI light curves.

\begin{deluxetable*}{ccclrrr}[ht]  \label{tbl:4TESS}
\centering
\tablewidth{6in}
\caption{Periodogram Performance for Four TESS Planet Candidates}
\tablehead{
\colhead{DTARPS-S} & \colhead{TIC} & \colhead{Name/Disp} & \colhead{Period} & \colhead{Depth} & \colhead{SNR} & \colhead{FAP} \\
\colhead{(1)} & \colhead{(2)} & \colhead{(3)} & \colhead{(4)} & \colhead{(5)} & \colhead{(6)} & \colhead{(7)}
}
\startdata
27~ & 22221375  & TOI~652.01 & 3.99 & 300~~ & \nodata & \nodata~~  \\
    &  BLS      & DTARPS-S~27 & 3.98387 & 322~~ & 17.4& $< 10^{-6}$ \\
    &  TCF      & CP& 3.98205 & 310~~ & 32.9& $< 10^{-6}$ \\
    & \\
98~ & 78669098  & \nodata & 0.57787 & 700~~ & \nodata & \nodata~~  \\
    &  BLS      & DTARPS-S~98 & 0.57764 & 477~~ & 24.2 & $< 10^{-6}$ \\
    &  TCF      & \nodata & 0.57778 & 530~~ & 63.7 & $< 10^{-6}$ \\
    & \\
103 & 89020549 & TOI~2336.01 & 2.11173 &1200~~ & \nodata & \nodata~~  \\
    &  BLS      & DTARPS-S~103 & 2.11076 & 1048~~ & 18.8 & $< 10^{-6}$ \\
    &  TCF      & CP & 2.11045 & 986~~ & 22.6 & $10^{-5}$ \\
    & \\
197 & 176685457 & TOI~1935.01 & 4.43427 & 6600~~ & \nodata & \nodata~~ \\
    &  BLS      & DTARPS-S~197 &  4.43031 & 4035~~ & 10.2 & $6 \times 10^{-4}$ \\
    &  TCF      & CP & 4.44109 & 3185~~ & 18.9 & $3 \times 10^{-3}$ \\
\enddata
\tablecomments{
The associated periodograms are shown in Figures~\ref{fig:real1}-\ref{fig:real4}. \\ 
Col.\ 1: DIAmante TESS AutoRegressive Planet Search South identifier from \citet{Melton23a}. \\ 
Col.\ 2: TESS Input Catalog identifier. \\
Col.\ 3: Names and Disposition. Row 1 gives the TOI designation from \url{https://tess.mit.edu/toi-releases/}. Row 2 gives the DTARPS-S designation from \citet{Melton23a} based on ARIMA + TCF analysis. Row 3 gives the disposition from \url{https://exoplanetarchive.ipac.caltech.edu} where CP = Confirmed Planet.
Col.\ 4. The orbital period from the NASA Exoplanet Archive (first line) and the two periodograms (second and third line).  \\
Col.\ 5: Depth in ppm.  \\
Col.\ 6: Peak Signal-to-Noise Ratio for the detrended periodogram.  \\
Col.\ 7: Extreme Value Theory False Alarm Probability for the standardized periodogram. }
\end{deluxetable*}


\section{Discussion}   \label{sec:discussion}

\subsection{Summary of the study}

This paper provides a detailed comparison of the Box-Least Squares (BLS) and Transit Comb Filter (TCF) periodograms to optimize the detection of small exoplanets from light curves with regular cadences as obtained from space-based surveys. The analysis has several steps:
\begin{enumerate}
    \item Tests are first conducted on simulated light curves with Gaussian white noise and autocorrelated noise by varying the number of transits, planet period, and transit duration (\S\ref{sec:sim}). The simulated light curves resemble single-sector TESS observations. The analysis is repeated for four observed TESS light curves with known small planetary transit signals (\S\ref{sec:TESS}).
    
    \item The light curves are detrended with a Gaussian Processes regression model prior to applying the BLS periodogram and with an ARIMA regression model prior to applying the TCF periodogram (\S\ref{sec:simulations}). Both remove long timescale trends but the latter also removed short-memory stochastic autocorrelation. 
    
    \item Two statistical metrics are applied to decide whether a periodogram peak represents a true planetary signal: a Signal-to-Noise Ratio (SNR) measured locally in the periodogram using a robust noise measure and a False Alarm Probability (FAP) based on extreme value theory (\S\ref{sec:per_snr}-\ref{sec:per_EVT}). In the latter case, a Generalized Extreme Value (GEV, \S\ref{sec:evt}) statistical model is fitted to the peaks of bootstrapped periodograms calculated using only a portion of the frequency range to alleviate high computational costs, as proposed by \citet{Suveges14}. 
    
    \item The sensitivity of a given periodogram is quantified with a Minimum Detection Depth (MDD) measure defined as the smallest statistically significant transit depth (\S\ref{sec:results_LCprop}). 
\end{enumerate} 

Our most noteworthy finding for both simulated and observed light curves is that TCF's periodogram peak shows a larger SNR than BLS in all experimented cases when the number of transits in the light curve is below $\sim$20 ($\sim$100) transits for Gaussian (autoregressive) noise (\S\ref{sec:results_LCprop}). {\it This demonstrates that the TCF periodogram following ARIMA modeling is more sensitive to small planets than BLS (following some local regression procedures like Gaussian Processes or spline fitting) using the SNR criterion for shorter duration light curves.} TCF shows no degradation in sensitivity to small planets as the number of transits in the light curve drops, even for only $2-3$ transits.

It may seem surprising that TCF outperforms BLS even for the pure Gaussian white noise simulations, as a least squares procedure gives a maximum likelihood estimator according to the Gauss-Markov Theorem. However, the theorem only applies to homoscedastic independent noise, while periodograms have heteroscedastic (and very non-Gaussian) power distributions, as seen in the histograms of periodogram power values. We discuss this issue below (\S\ref{sec:BLS_poor-performance}). 

We find that the FAP and SNR metrics lead to opposing conclusions when pure Gaussian white noise is simulated in light curves but similar conclusions for autocorrelated noise. Experiments on Gaussian white noise suggest that BLS is slightly more sensitive than TCF using the FAP criterion, whereas TCF is more sensitive using the SNR criterion. For the FAP metric and light curves with autocorrelated noise (upper right panel of Figure \ref{fig:ntransitsBLSTCF}), the sensitivities of both periodograms using the FAP criterion are degraded compared to the case of pure Gaussian white noise light curves, particularly when few transits are present. However, this is not a critical problem, as the SNR metric is remarkably unaffected by the number of transits for the TCF periodogram.  

The four TESS light curves analyzed in \S\ref{sec:TESS} have milder autocorrelation than in our simulations but are significant enough to distinguish it from white Gaussian noise. All four planets readily passed our significance criteria (periodogram peak $\textrm{FAP} < 0.01$ or $\textrm{SNR} > 6$) for both BLS and TCF with the correct orbital periods.  

Altogether, the most sensitive approach to small planet discovery is using the Transit Comb Filter periodogram preceded by ARIMA modeling of the light curve and a robust signal-to-noise ratio metric. These findings explain why TCF had high sensitivity to small planets in previous studies: \citet{Caceres19b} applied the ARIMA + TCF procedure to $\sim$150,000 Kepler light curves and reported 97 Earth- and Mars-sized planetary candidates from the 4-year Kepler data, and \citet[][Figure~16]{Melton23c}  applied the procedure to $\sim$1 million TESS light curves and reported hundreds of candidate planets substantially smaller than Confirmed Planets in the Year 1 TESS survey. It is also being used in Pellegrino et al. (in preparation) for Year 2 TESS data. As investigated by our experiments here, an advantage of SNR over FAP is that while the latter tends to fluctuate non-trivially across multiple simulated noise realizations (\S\ref{sec:resultsSetup}), the former was found to be relatively more stable. 

The problems of heteroscedasticity and trends in BLS periodograms may be partly ameliorated by different choices of detrenders. For example, \citet{Hippke2019Detrending} recommend a robust time-windowed filter procedure. However, as Figure~\ref{fig:ntransitsBLSTCF} shows, the TCF periodogram provides more sensitivity to small planets (using the preferred metric, SNR) than BLS even for Gaussian white noise. This suggests that TCF is preferred regardless of the detrending procedure used before BLS.

Another contribution of this paper is to present a broad approach for comparing periodograms. Previous studies such as of \citet{Graham13} performed a comparative study of periodograms using relevant metrics, however, a concrete methodology was not sought. While the application of our periodogram comparison approach has been limited to two periodograms for transiting planet detection, the approach is easily extensible to compare any set of periodograms having similar aims (e.g., comparing the Fourier Schuster, Lomb-Scargle, Phase Dispersion Minimization, BLS and TCF periodograms) since the approach is agnostic to the type of periodogram. When applied to periodograms where significance is not the primary metric, the MDD criterion can be changed to any other criterion relevant to the task, and the comparison approach can still be used by changing the x-axis in Figure~\ref{fig:ntransitsBLSTCF} from the number of transits to some other parameter of the light curve. The R code we developed is publicly available\footnote{ See the Software notes at end of paper.} and can be used to extend the comparison study for different applications.

\begin{figure*}[ht]
\centering
\includegraphics[keepaspectratio,width=0.44\linewidth]{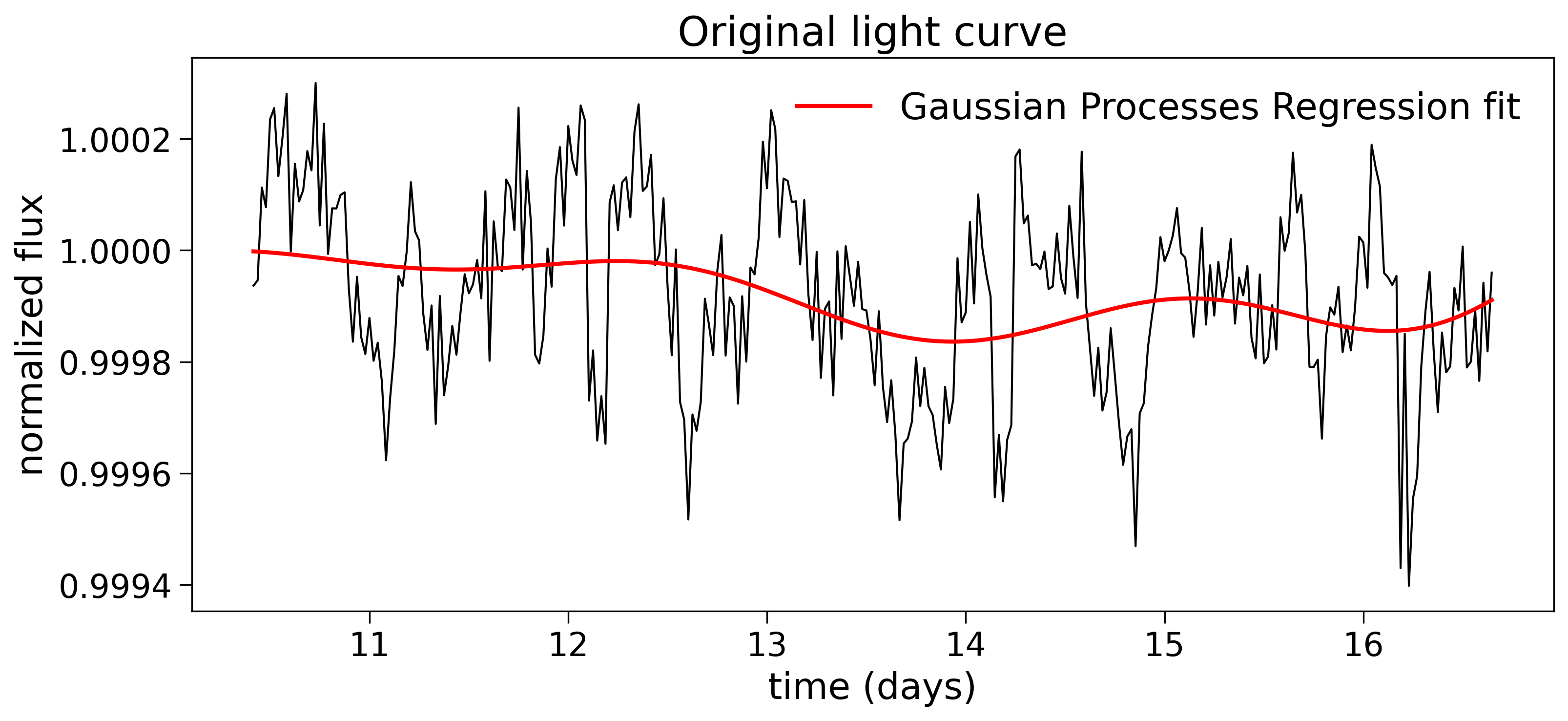}
\includegraphics[keepaspectratio,width=0.44\linewidth]{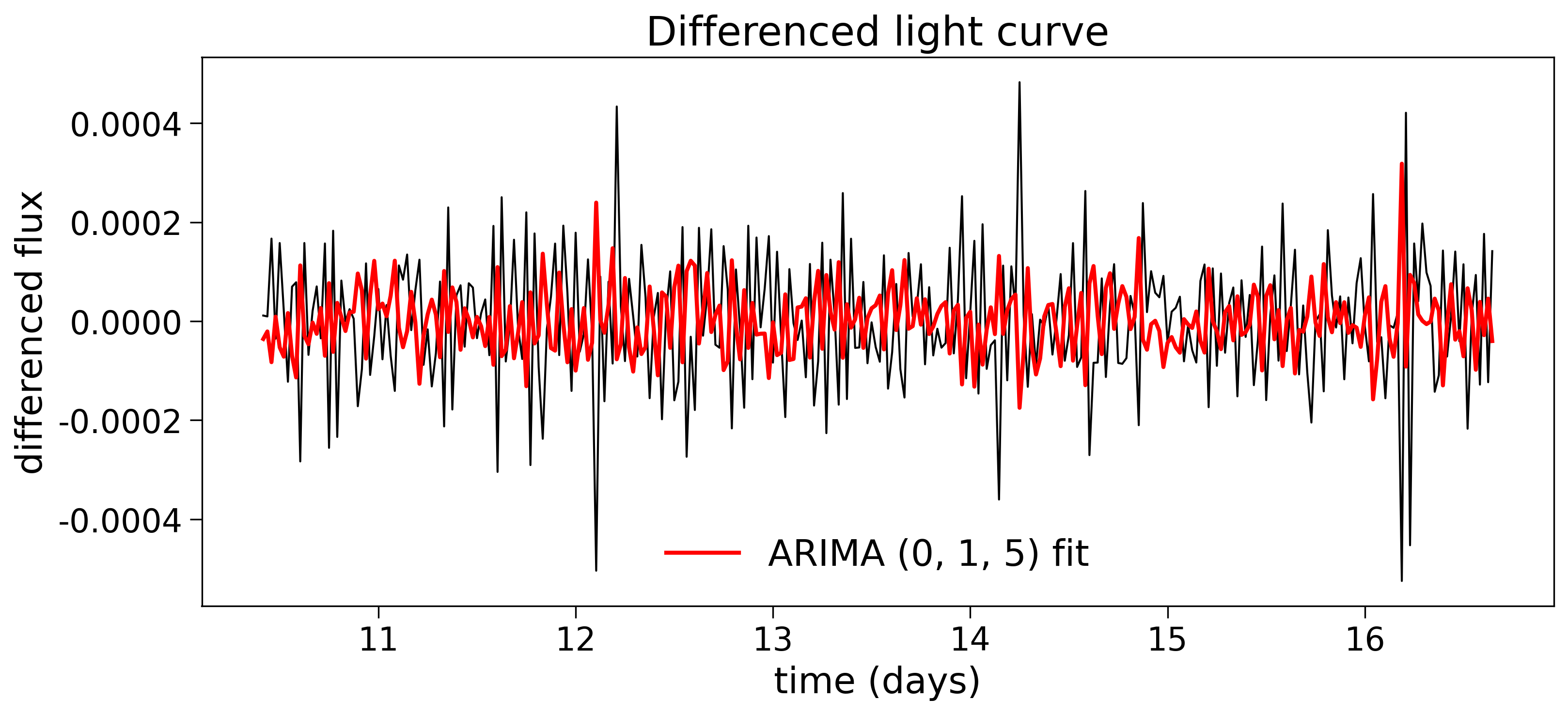} \\

\includegraphics[keepaspectratio,width=0.44\linewidth]{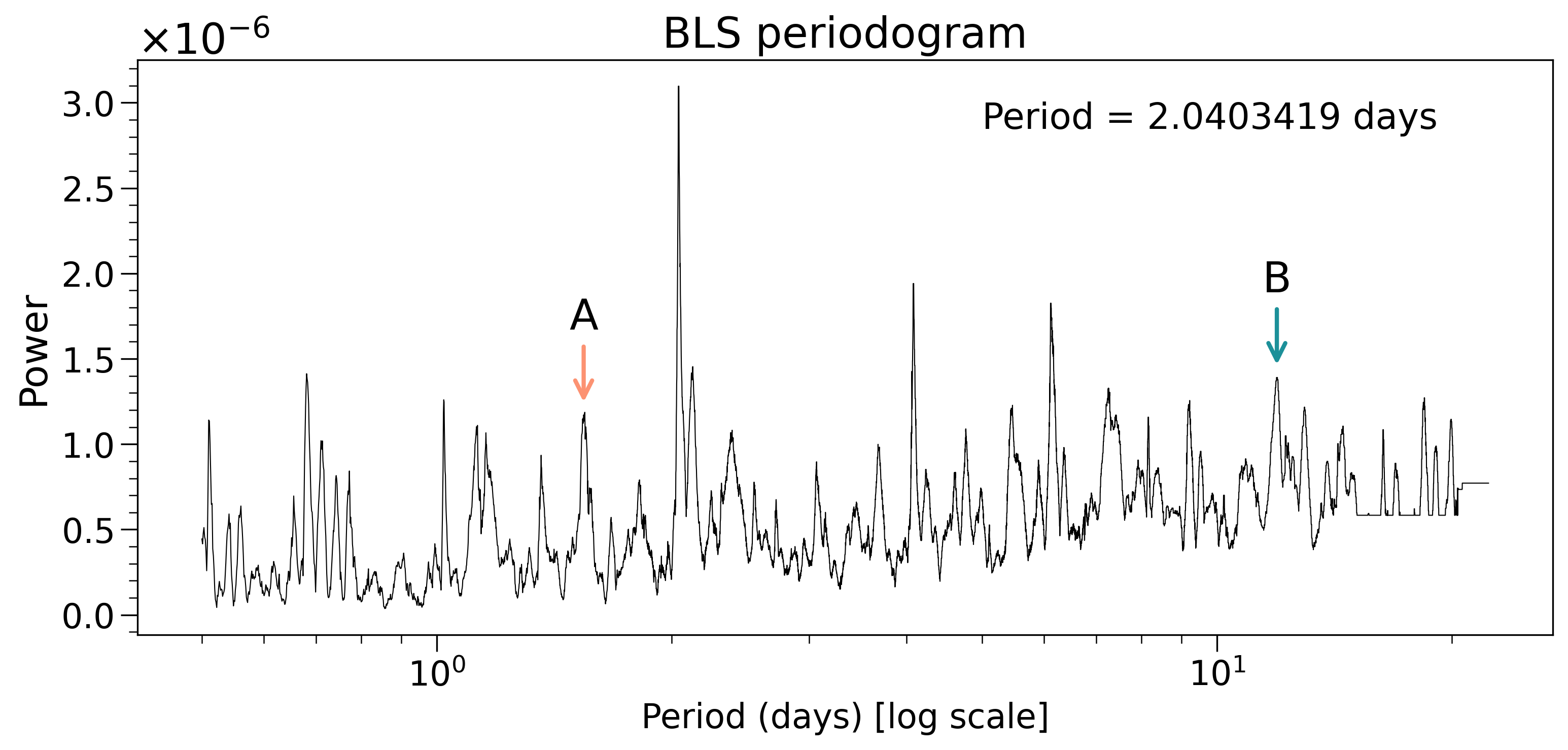}
\includegraphics[keepaspectratio,width=0.44\linewidth]{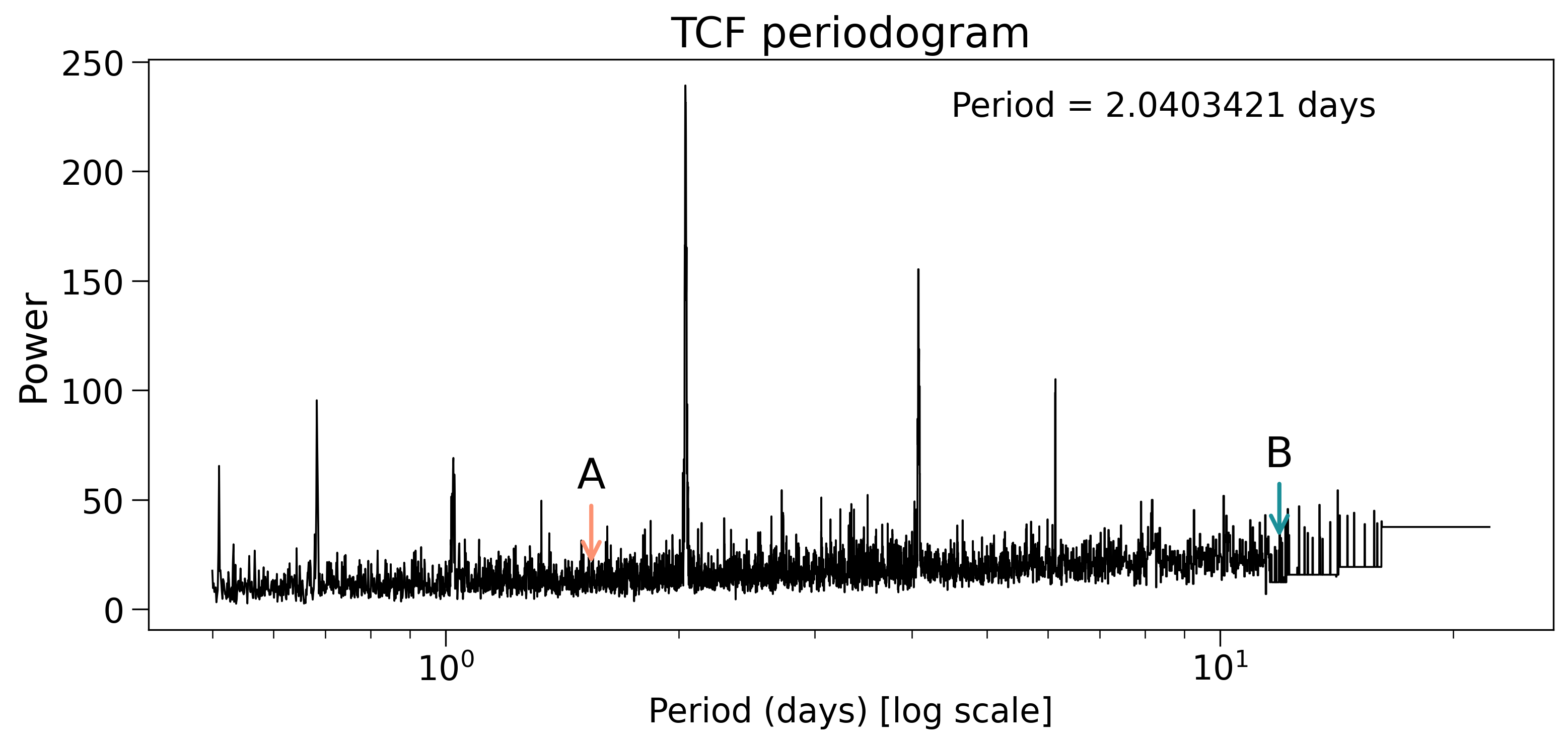} \\

\includegraphics[keepaspectratio,width=0.44\linewidth]{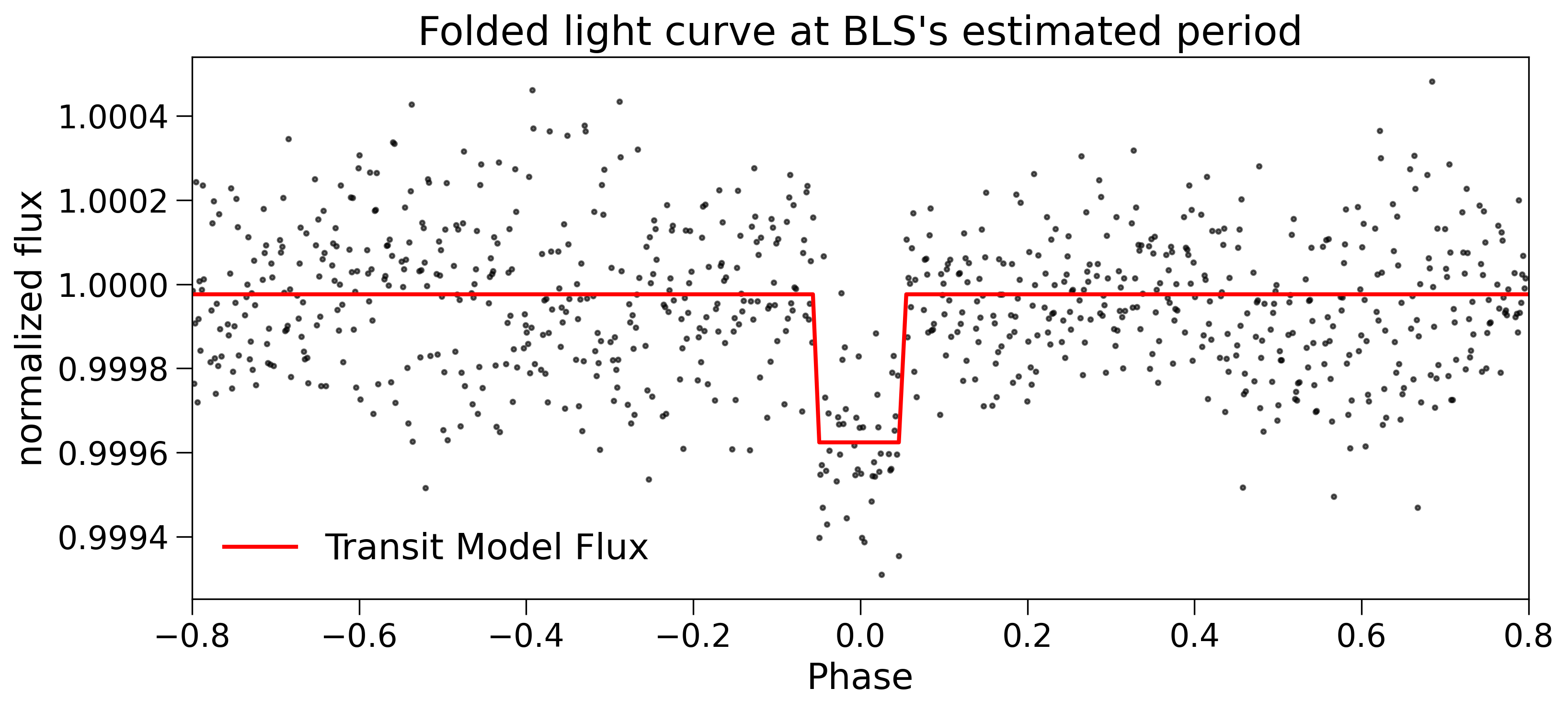}
\includegraphics[keepaspectratio,width=0.44\linewidth]{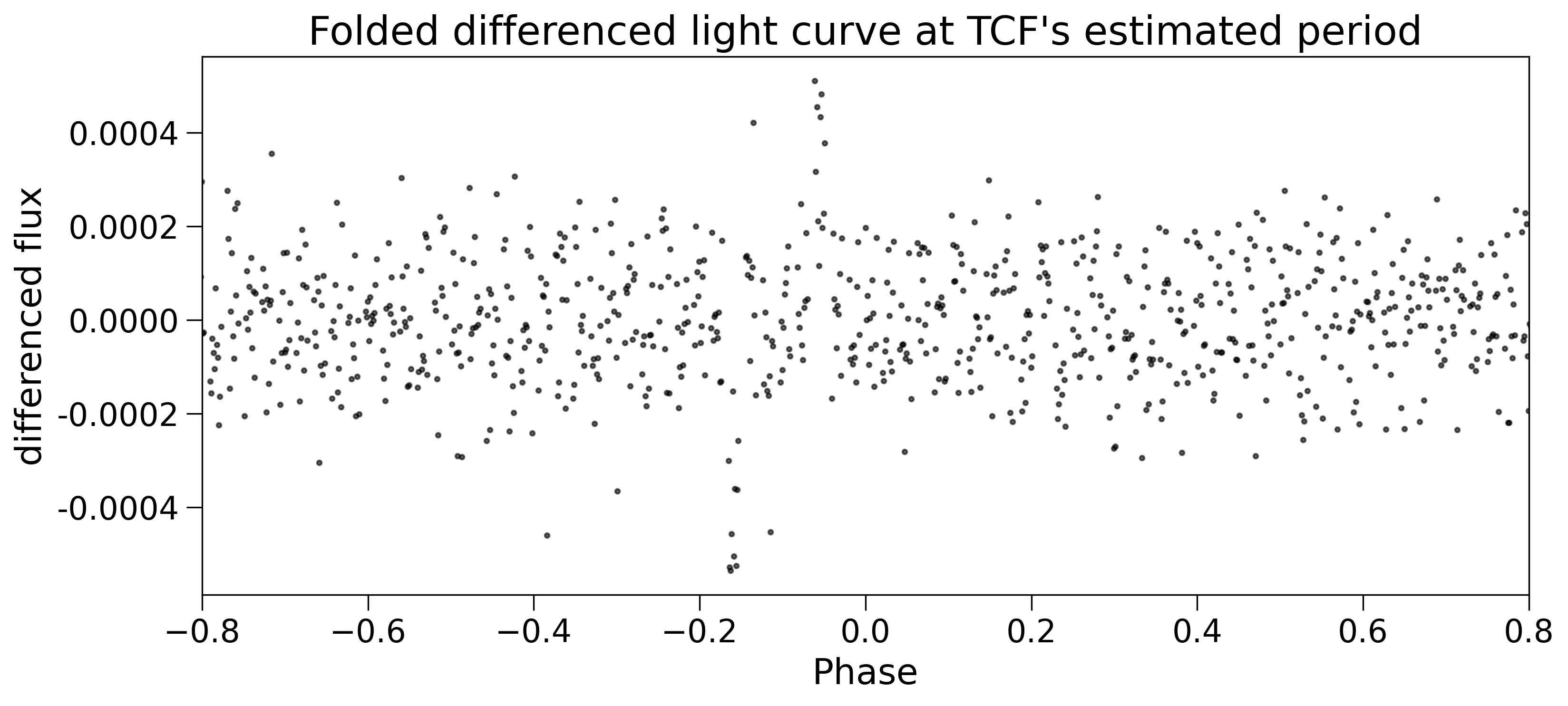} \\

\includegraphics[keepaspectratio,width=0.44\linewidth]{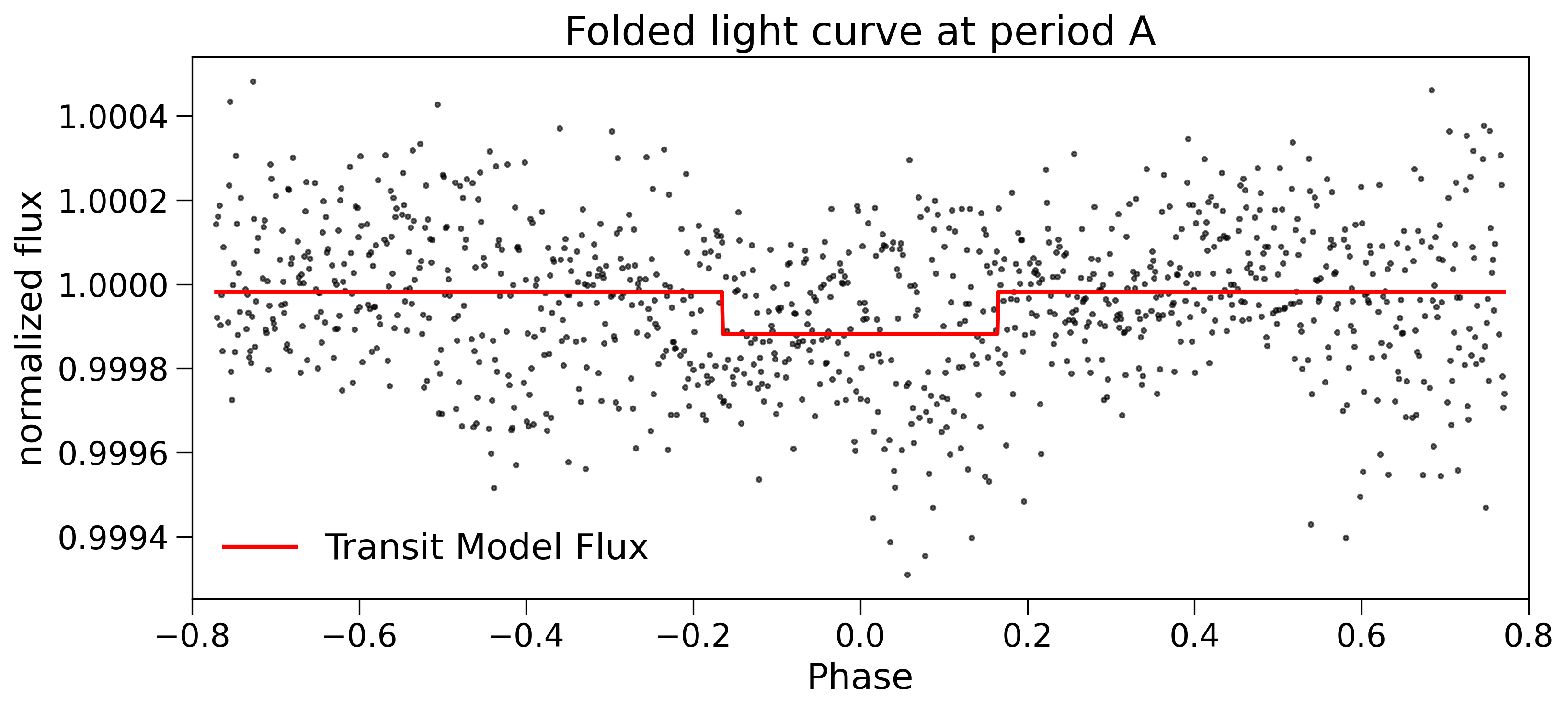}
\includegraphics[keepaspectratio,width=0.44\linewidth]{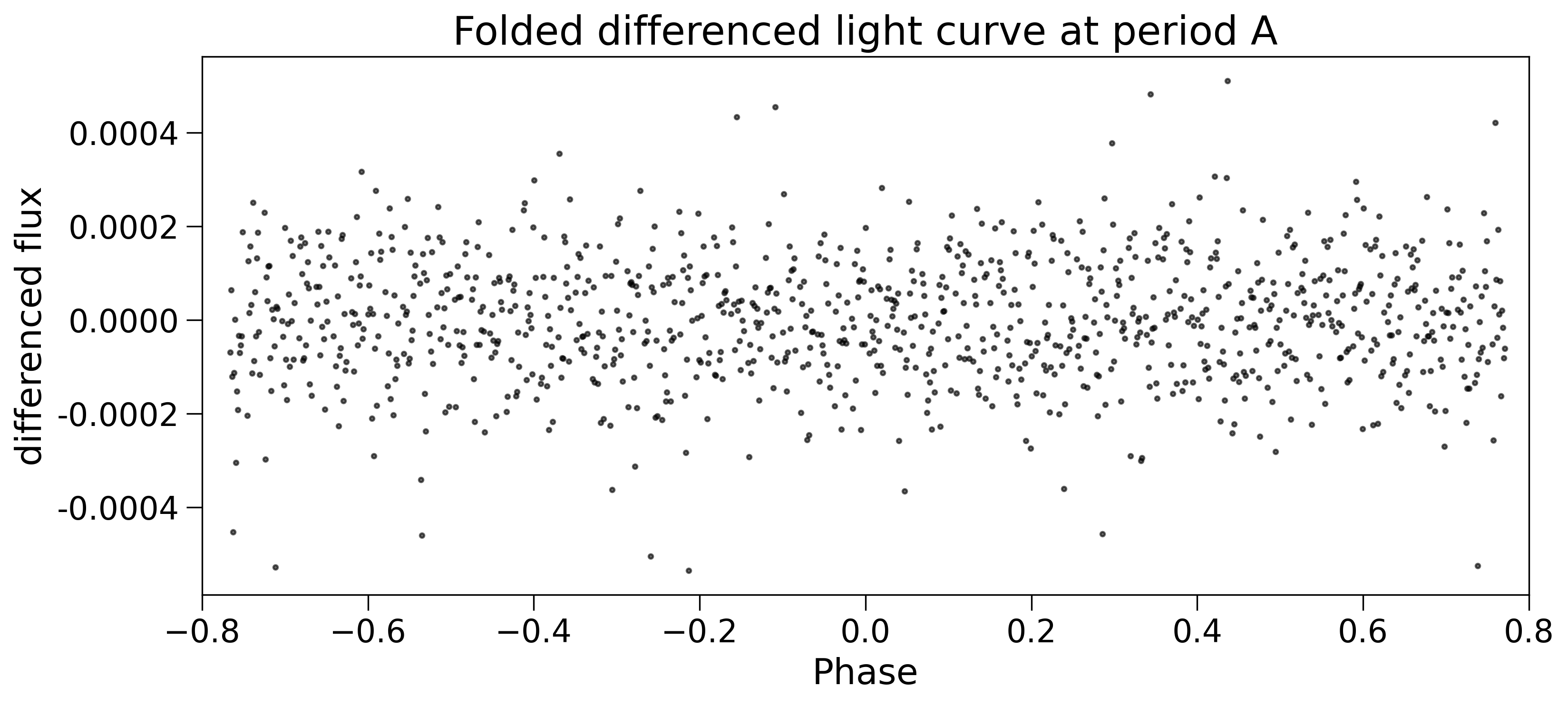} \\

\includegraphics[keepaspectratio,width=0.44\linewidth]{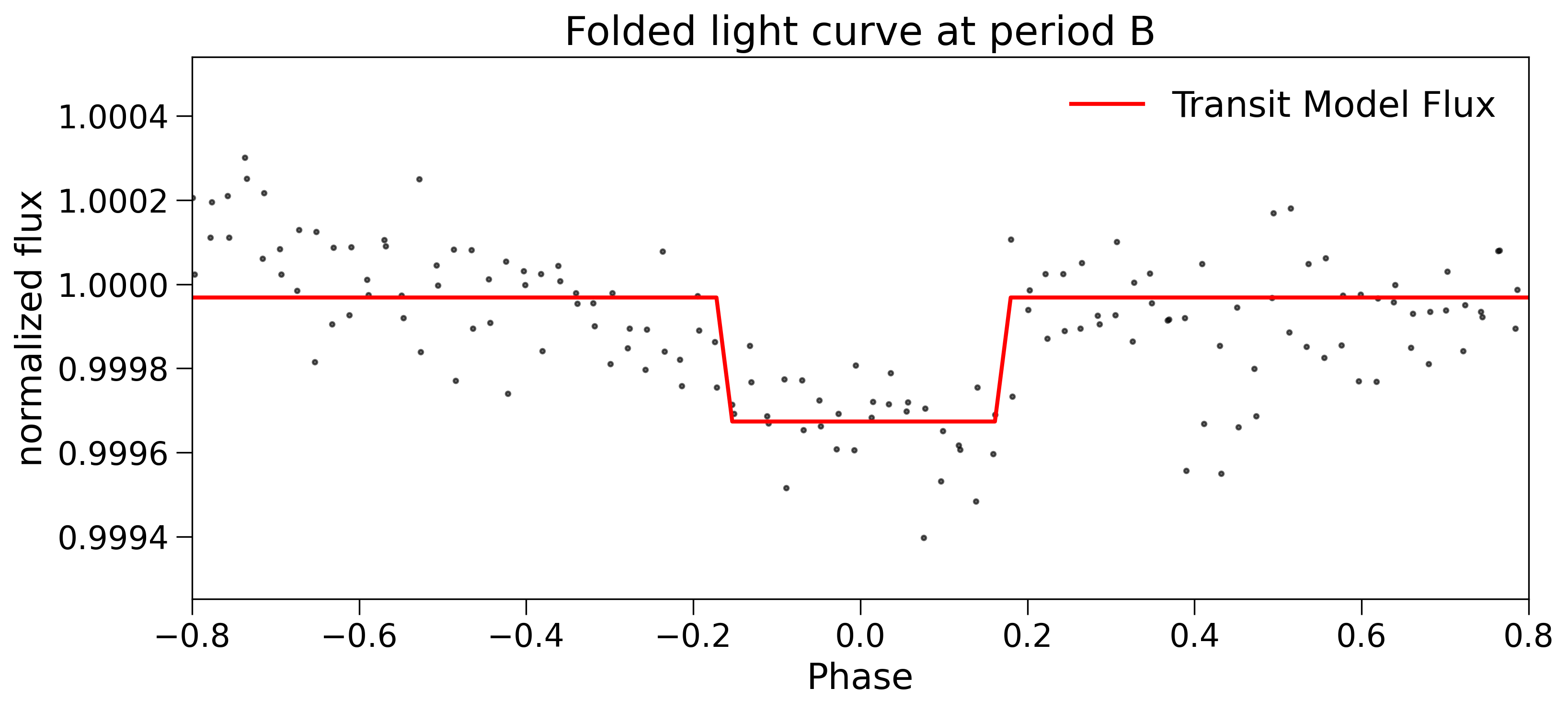}
\includegraphics[keepaspectratio,width=0.44\linewidth]{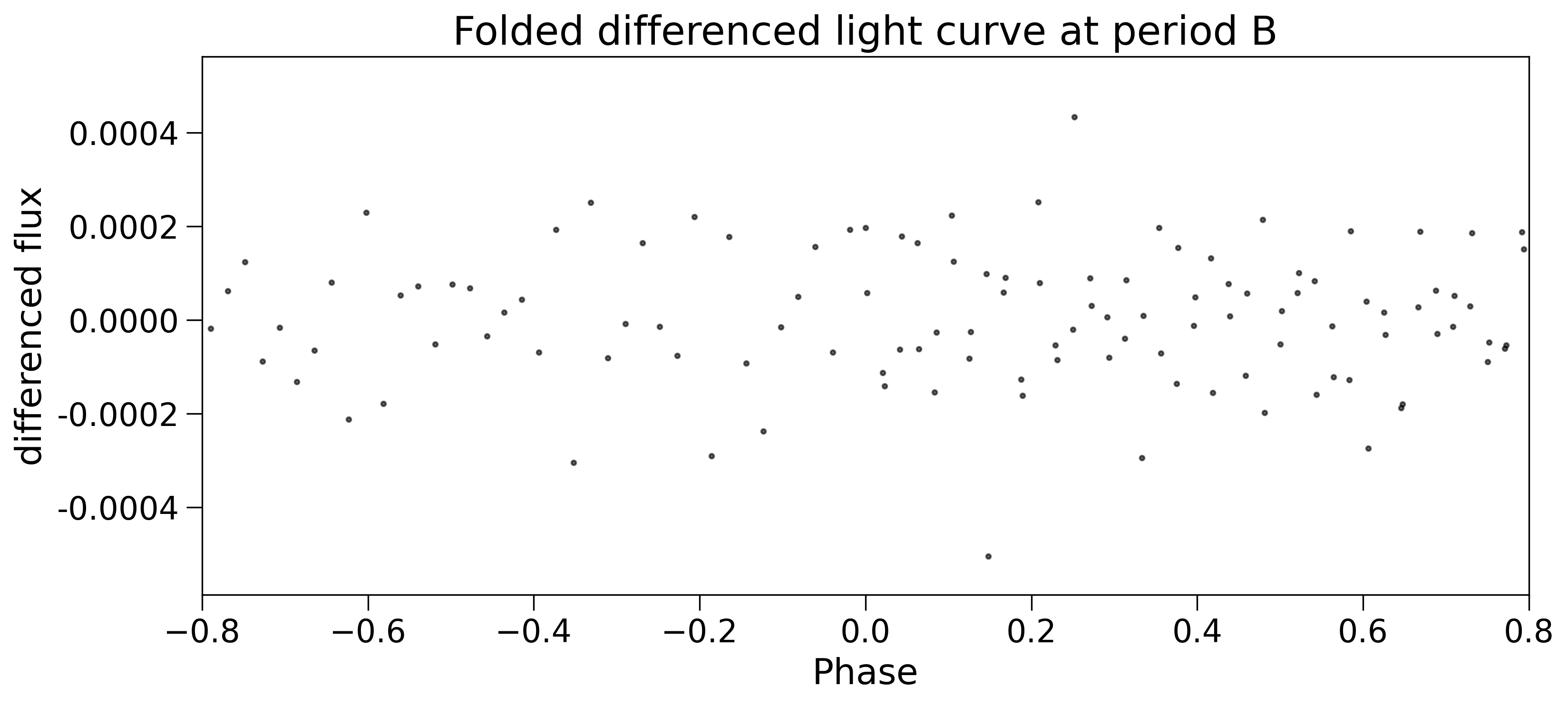} \\

\caption{Illustration of BLS (left panels) and TCF (right panels) analysis of a simulated light curve with a P=2.00 day transiting planet superposed on autocorrelated noise.  {\it First row:} Zoom of several days of the light curve with the Gaussian Processes Regression fit overlayed (left) and the same light curve after differencing with the ARMA fit overlayed (right).  {\it Second row:} Periodograms of the entire simulated light curve with two BLS false peaks `A’ and `B’ marked.  {\it Third row:} Folded light curves for the true transit period. {\it Fourth and fifth rows:} Folded light curves of the original (left) and differenced (right) data for false spectral peaks A and B. }
\label{fig:whyTCFPerformsSoWell}

\end{figure*}

\subsection{Why does BLS perform so poorly?}
\label{sec:BLS_poor-performance}

Ideally, both simulated noise models (white Gaussian and autoregressive) should have resulted in similar conclusions since BLS and TCF are preceded by detrending procedures that should have removed any correlation structure from the light curve. However, our findings show that BLS (preceded by Gaussian Processes regression detrending) is less sensitive to small planets than TCF (preceded by ARIMA regression detrending). We investigate the causes of this difference in sensitivity with a detailed examination of the inner workings of both algorithms.

Figure~\ref{fig:whyTCFPerformsSoWell} shows an example of a simulated planet with period = 2.00 days, transit duration = 2.00 hrs, depth = 0.04\%, ten transits in the light curve, and autocorrelated noise with ARMA (3, 3) from equation \ref{eqn:arma}\footnote{We used the \texttt{Lightkurve} Python package \citep{2018ascl.soft12013L}, version 2.3.0 for generating the folded light curves, showing the BLS transit models, and for the corresponding plots.}. The depth is chosen sufficiently large so that BLS and TCF peaks are significant using the FAP and SNR criteria.

The top-left plot shows the original light curve with the Gaussian Processes Regression fit overlayed; the top-right plot shows the differenced light curve with the ARMA fit. This Gaussian Process fit clearly misses most of the short-memory structure, although different kernel hyperparameters might do a better job. The second row shows the corresponding BLS and TCF periodograms. The same set of test periods is used; here, we omit the Gaussian Processing detrending to better highlight BLS's characteristics in the case of short-memory autocorrelation. The BLS periodogram exhibits higher and spikier noise and a stronger rising trend with period than the TCF periodogram, as seen earlier in Figures \ref{fig:simGauss1}-\ref{fig:simARMA2} and Figures \ref{fig:real1}-\ref{fig:real4}. 

The fourth and fifth rows of Figure~\ref{fig:whyTCFPerformsSoWell} examine two false peaks marked by labels ``A" and ``B" corresponding to shorter and longer periods in the periodograms, respectively. The folded light curves for the correct period clearly show a box-like transit for BLS and a double spike for TCF (third row). Since periods A and B are not the true period, their light curves should not possess box-like shapes for BLS and double-spike for TCF. However, the fourth and fifth rows of Figure~\ref{fig:whyTCFPerformsSoWell} illustrate that model fitted by BLS tends to capture chance alignments of outliers or autocorrelated ripples even when the folded light curve possesses no transits. While TCF can also match a double spike pattern at non-transit periods, it is used only when the autocorrelation is removed via ARIMA modeling. TCF only considers extreme points in the differenced light curve that proves to be less susceptible to random chance alignments. 

We infer from Figure~\ref{fig:whyTCFPerformsSoWell} that BLS periodograms are often noisier than TCF periodograms, particularly for light curves with short-memory autocorrelation remaining after inadequate detrending, because chance alignments of the structure can easily mimic box-like shapes. On the other hand, the double-spike structure matched by the TCF algorithms is difficult to reproduce by autocorrelation alone, so the TCF periodogram noise is better behaved. The trend of increasing BLS power as the period increases for all periods without true periodic signals can be attributed to the weaker dilution of autocorrelated patterns in the folded light curves. This can be seen by comparing the higher depth of the false box in the fifth row compared to the false box in the fourth row.

It is reasonable that the BLS algorithm (with a standard detrender like Gaussian Processes regression) produces more noise than the TCF algorithm (with an ARIMA detrender designed to remove stochastic short-memory autocorrelation) when the detrended light curve has significant autocorrelation. However, we are surprised that TCF can outperform BLS with the signal-to-noise ratio metric when the light curve is mostly white Gaussian noise (Figures~\ref{fig:simGauss1}-\ref{fig:simGauss2} and the lower-left panel of Figure~\ref{fig:ntransitsBLSTCF}).  We believe the TCF algorithm is more stable because it seeks a distinctive double-spike pattern of a few brightness outliers, which are sparse and unlikely to be aligned in light curves folded with random periods.

\subsection{Advice for transit searches}

Figures~\ref{fig:ntransitsBLSTCF}-\ref{fig:periodAndDurationCompare} have shown that a combination of TCF with the SNR metric achieves excellent sensitivity to small transiting planets for a wide range of transit periods and durations, whereas the other combinations of detection methods (TCF-FAP, BLS-FAP, and BLS-SNR) were relatively less sensitive. 

Based on this result, we recommend the following procedure for small planet detection from a transit survey:
\begin{description}
    \item [Detrend the light curve] Compute the nonparametric autocorrelation function of the light curve. In the case of irregularly-spaced light curves, the traditional ACF estimation as used in this study may not be defined, so some modifications are needed (see, e.g., \citealt{1989ApJ...343..874S}; \citealt{2005ASPC..334..659A}). If the ACF deviates significantly from white noise (based on the Durbin-Watson and Ljung-Box hypothesis tests), then fit the best ARIMA model with complexity determined by the Akaike Information Criterion. A single differencing step should be used so that box-shaped transits are converted to double-spike patterns\footnote{
    We recommend using the $auto.arima$ function in the $forecast$ CRAN package within the R statistical software environment described in the volume by \citet{Hyndman21}. The $forecast$ package, downloaded $\sim 8000$ times per day for many purposes, is highly capable and reliable. Details on the $forecast$ package can be found at \url{https://pkg.robjhyndman.com/forecast/} and \url{https://cran.r-project.org/web/packages/forecast}. A complete version of our pipeline can be found at \url{https://github.com/Yash-10/Periodogram-Comparison-Optimize-Planet-Detection}. The specific command sequence we recommend is: \\   
    diff.lc ~=~ c(NA, diff(lightcurve.values) \# Single differencing operation \\
    ARIMA.fit ~=~ auto.arima(diff.lc, stepwise=FALSE, approximation=FALSE, seasonal=FALSE, max.p=5, max.q=5, max.d=0) \# Find best fit up to complexity ARIMA(5,1,5). 
    }. 
    Repeat the autocorrelation function of the residuals to see whether they approach uncorrelated white noise. 

    \item [Compute the TCF periodogram] This search for periodic double-spike patterns is calculated on the ARIMA residuals for a chosen range of periods with an oversampling of trial periods so that true spectral peaks are not missed. The periodogram is standardized in a robust fashion (footnote 4).
    
    \item [Identify the best trial transit period] The robust SNR metric in equation (\ref{eqn:SNR}) is calculated for each trial period of the standardized TCF periodogram. The highest SNR peak represents the best possibility of a transiting planet detection.  
    
    \item [Reduce statistical False Alarms] This is a multifaceted step that can include: an examination of the periodogram for noise peaks comparable to the best peak, examination of folded light curves for patterns inconsistent with true periodic photometric dips, and construction of simulated light curves of the suspected transit. The periodograms of the simulations should be examined for alias structure and MDD sensitivity for comparison with the observed periodogram. 
    
    \item [Reduce astronomical False Positives]   These steps $-$ discussed by \citet{Guerrero21}, \citet{Melton23b} and others mentioned in \S\ref{sec:intro} $-$ lie beyond the scope of our discussion here. 
\end{description}

\begin{figure*}
    \centering
    \includegraphics[keepaspectratio,width=0.8\linewidth]{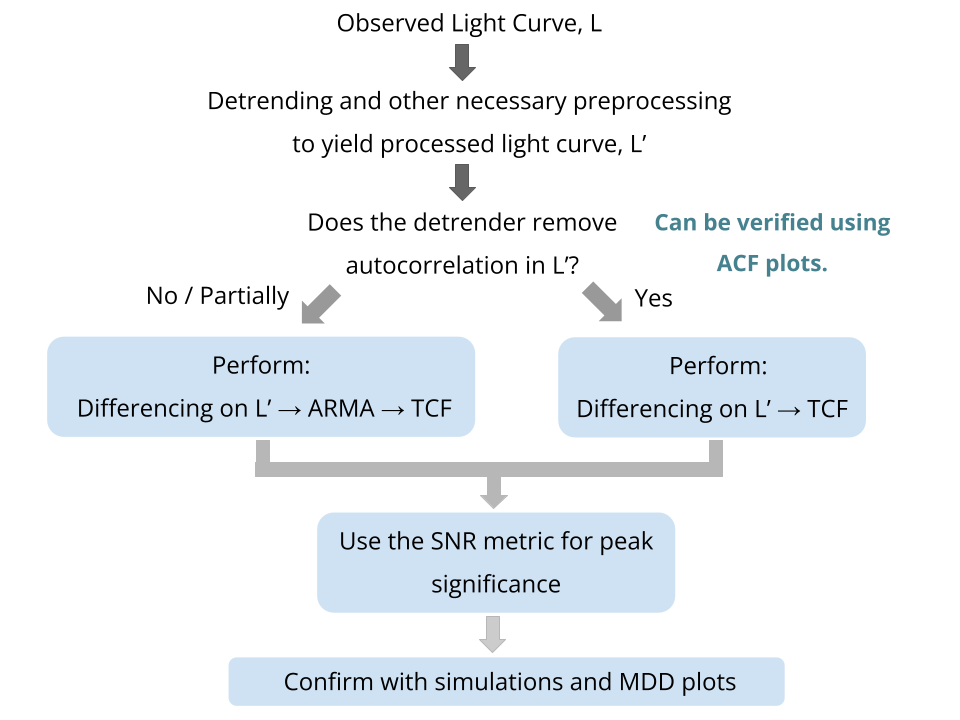}
    \caption{Decision tree outlining general suggestions for selecting a periodogram algorithm based on certain conditions.}
    \label{fig:decisionTree}
\end{figure*}

These suggestions are summarized in Figure~\ref{fig:decisionTree}. However, it should not be considered the ultimate guide. In particular, we recommend making new versions of the MDD plots similar to Figure~\ref{fig:ntransitsBLSTCF} based on the characteristics of the transit survey (noise level, non-Gaussian behaviors, observation duration, etc.) under study.

An additional analysis procedure can be considered to help adjudicate the reality of a small planet transit signal. First, \citet{Hippke19} shows that incorporating astrophysical knowledge can improve sensitivity to small planets. Here the square box shape transit of BLS and TCF is curved due to stellar limb darkening. The limb darkening shape depends on the star's effective temperature and surface gravity, which can be inferred from {\it Gaia} photometry and astrometry. This curved transit shape could be incorporated into a BLS algorithm or the TLS algorithm of \citet{Hippke19} {(see also \texttt{batman}; \citealt{Kreidberg15})}. Second, after a tentative periodicity has been identified from the TCF periodogram after ARIMA detrending, one might fit an ARIMAX model to the light curve that incorporates a deterministic box-shaped periodic transit component with the autoregressive component. This gives a new estimate of the transit depth with uncertainty based on the model Fisher information matrix. The SNR of the resulting transit depth can be beneficial for subsequent analyses. See \citet{Caceres19a} (\S3.3) for more details.

ARIMA + TCF and TLS are complementary approaches: while TLS improves the sensitivity of BLS by adding astrophysical insights, TCF is more sensitive than BLS because of an effective treatment of stellar autocorrelated noise, resulting in reduced periodogram noise. The standard ARIMA + TCF procedure could be followed by TLS, a more refined periodogram incorporating limb darkening, and a self-consistent pattern of transit ingress and duration for the particular stellar and planetary inferred parameters from the standard procedure.

\section{Conclusion}\label{sec:conclusion}

This paper cautions the reader about weaknesses in the sensitivity of the commonly used BLS periodogram for detecting small planets. Problems are exacerbated when autocorrelation persists in the light curve, even after detrending. BLS has the unfortunate characteristic of fitting boxes to noise for both autocorrelated and Gaussian white noise. This results in spurious peaks and poor statistical properties (heteroscedasticity and trends) discussed by \citealt{Ofir14}. These factors inhibit the detection of small planets using BLS, as previously noted by studies such as that of \citet{Hippke19}.

The main achievement of this paper is explaining the advantages of ARIMA + TCF procedure from the AutoRegressive Planet Search project \citep{Caceres19a} has improved performance over standard detrenders with BLS periodogram. ARIMA, a widely used method for modeling stochastic autocorrelated time series since the 1970s \citep{Box15}, detrends (in most cases) both longer-term trends and stationary short-memory autoregressive behaviors in the light curve. It is followed by the Transit Comb Filter that matches the sharp ingress and egress spikes in the ARIMA residuals for a trial period, providing the cadence is well-matches to ingress timescales. 

The ARIMA + TCF pipeline proves to be remarkably effective, and we show that much of the advantage emerges from the TCF periodogram. It has a lower noise with weaker spurious peaks than the BLS periodogram, even for time series with white Gaussian noise. TCF also scales slightly better in terms of computation time compared to BLS. 

We further find that the commonly used signal-to-noise ratio is the preferred metric for optimizing small exoplanet detection compared to a False Alarm Probability based on extreme value theory. The latter is now often used with the Lomb-Scargle periodogram \citep{Baluev08, Suveges14}. Section 5 of \citet{Caceres19b} and Figure~16 of \citet{Melton23c} show that transit planet candidates derived from the ARIMA + TCF combination  are often smaller than confirmed planets derived from BLS-based procedures, and thus the findings of this paper. Our study explains some of the problems with the BLS periodogram described by \citet{Ofir14}. Section \ref{sec:BLS_poor-performance} elucidates why the BLS periodogram often has unnecessarily high noise and spurious peaks where no periodicity is present.

Our study also shows that analysis of simulations can be effective for evaluating the periodogram peak significance on observational data, particularly when complicated conditions (such as missing or autocorrelated data in observations) are present. Analysis of simulations can be used to compare any combination of periodograms. Simulations were also recommended by \citet{VanderPlas18} in his discussion of the significance of Lomb-Scargle periodogram peaks.  

The results emphasize that effective detrending algorithms are vital to improving periodogram sensitivity. We use the classical sequence of differencing and ARMA modeling. However, combining spline or Gaussian Processes regression for longer-term trends with ARMA for short-memory stochastic autocorrelation is an intriguing possibility that may be beneficial for BLS. However, our preliminary experiments (not reported here) suggest that BLS performance remains inferior. 

Any procedure seeking efficient small planet detection must avoid fitting the light curve too well so the planetary signal is absorbed into the detrending model.  The ARIMA method of differencing and fitting an ARMA model leaves most of the planetary signal untouched since, once the light curve is differenced, the ARMA model does not fit the double spike since most of the in-transit points are removed. It allows the use of the TCF algorithm that gives a well-behaved periodogram with more homoscedastic noise and reduced spurious peaks. Novel methods for removing stellar trends have been proposed (such as \citealt{irf} and \citet{Smith12}) and could be studied in combination with periodogram analyses for detecting small transiting planets in upcoming exoplanet missions. The incorporation of astrophysically motivated transit shapes can be combined with ARIMA + TCF to improve transit survey sensitivity further.

In conclusion, the ARIMA-TCF procedure instead of  local regression detrenders and BLS can significantly improve the sensitivity of space-based photometric surveys for detecting small planets. The degree of improvement depends on several factors, such as the number of transits and the noise characteristics in the light curve. Prospective applications of our study include existing COROT, Kepler, K2, and TESS datasets and forthcoming data products of ESA's PLATO mission and NASA's Roman Space Telescope. 

\begin{acknowledgments}

The AutoRegressive Planet Search project at Penn State has been funded by NASA grant 80NSSC17K0122 and NSF grant AST-1614690. S. Saha would like to thank the Science and Engineering research Board (SERB), SERB SURE (SUR/2022/001965), SERB CRG (CRG/2023/003210), Department of Science and Technology, Government of India (EMR/2016/005687) and APPCAIR, BITS Pilani K.K. Birla Goa campus for supporting this research.  We are grateful for the efforts of Elizabeth Melton (Rose-Hulman Institute of Technology) and Andrew Pellegrino (Penn State) for their substantial contributions to this project.  

This research has made use of the NASA Exoplanet Archive, which is operated by the California Institute of Technology, under contract with the National Aeronautics and Space Administration under the Exoplanet Exploration Program \citep{https://doi.org/10.26133/nea12}. It also benefited from  Lightkurve, a Python package for Kepler and TESS data analysis \citep{2018ascl.soft12013L}. The research used the online Kaggle platform for running a few long-running computations.
\end{acknowledgments}

\facilities{TESS \citep{Ricker15}} 

\software{BLS (Box-Least Squares periodogram)  \citep{2016ascl.soft07008K}; 
forecast \citep{Hyndman23}; 
microbenchmark \citep{microbenchmark}; 
PCOSTPD (Periodogram Comparison for Optimizing Small Transiting Planet Detection) \citep{2023ascl.soft09011G}; 
R \citep{RCoreTeam22}; 
TCF (Transit Comb Filter periodogram) \citep{2022ascl.soft06002C}. 
}
Our code implementation for estimating the statistical significance of periodogram peaks using the R programming language, with associated documentation and Jupyter notebook tutorial, is available online at three locations: at Zenodo (\doi{10.5281/zenodo.8431740}, frozen version), at the Astrophysics Source Code Library \citep{2023ascl.soft09011G}. and at Github (\url{https://github.com/Yash-10/Periodogram-Comparison-Optimize-Planet-Detection}, dynamic version).

\newpage
\appendix

\section{Periodograms and Extreme Value Theory}
\label{sec:evt}

Periodograms are a common way to search for periodicities in astronomical time series. They evaluate the value of a test statistic at chosen independent periods or frequencies \citep{VanderPlas18, Giertych22}. A candidate periodicity corresponds to a period where a measure of periodogram power is maximum and lies well above the periodogram noise.  

Evaluating the statistical and scientific significance of a trial periodicity is tricky. Periodogram power values may not follow a well-defined distribution, they may exhibit trends and heteroscedasticity with the period, and strong spurious (False Alarms) \citep{Brown03} or alias peaks may be present \citep{Frescura08, Ofir14, VanderPlas18}. A commonly used quantity to characterize the significance of the peak of a periodogram is the False-alarm Probability (FAP), the probability that the maximum value of the periodogram or higher is observed under the null hypothesis $H_0$ that the observed time series does not contain a periodic component. The alternative hypothesis $H_1$ is that a deterministic periodic signal is present. Typically, one chooses a significance level, such as $\alpha = 0.01$, and declares the peak significant, hence claiming detection of a transiting planet if $\textrm{FAP} < \alpha$.

A straightforward attempt to estimate the distribution of maximum\footnote{In some situations, it is not clear whether the extreme value is part of the stochastic process generating most of the data or whether it is a contaminant from another population. Since the periodogram power values are calculated in a consistent fashion from a single dataset, their peak values can not be extraneous contaminants and can be legitimately evaluated using extreme value theory.} $M_n = \textrm{max}\{X_1, X_2, \dots, X_n\}$,  where $X_1, X_2, \dots, X_n$ are independent and identically distributed (i.i.d) random variables, is
\begin{equation}
    \begin{aligned}
    P(M_n \leq z) & = P(X_1 \leq z, X_2 \leq z, \dots, X_n \leq z)\\
    & = P(X_1 \leq z) \times P(X_2 \leq z) \dots \times P(X_n \leq z)\\
    & = \{F(z)\}^n
    \end{aligned}
\end{equation}
where $F(z)$ is the cumulative distribution of the periodogram under the null hypothesis. In this case, the FAP can be written as 
\begin{equation}
\textrm{FAP} = 1 - \{F(z)\}^n.
\end{equation}

Such a formulation has been traditionally used in several studies (see, e.g., \citealt{Czerny98} for a theoretical background and summary). However, several difficulties arise in applying this FAP formula in real applications \citep[see discussion in][]{Suveges14}:
\begin{enumerate}

\item The assumption of i.i.d frequencies used for computing the periodogram does not hold in practice due to an irregular cadence or the oversampling of the periodogram needed to capture accurate peak values.

\item  It is not straightforward to evaluate $n$, the number of independent frequencies. This problem is widely discussed \citep[e.g.][]{Horne86, Frescura07, Frescura08, Koen10, VanderPlas18}. Generally, $n$ is interpreted as the ``effective" number of independent frequencies; however, its estimates are only an approximation even for evenly-spaced datasets \citep{Baluev08}.

\item The distribution function, $F(z)$, is generally unknown or known only approximately. While techniques to estimate $F$ from the sample exist, small errors in $F$'s estimation can cause large errors in $F(z)^n$. For example, the small difference between the observed light curve variance and its population variance can produce major errors in FAPs \citep{Koen90}.  

\item Consider the behavior of $F(z)^n$ when $n \to \infty$: since $0 \leq F(z) \leq 1$, $F(z)^n \to 0$ as $n \to \infty$, $\forall z < z_+$, where $z_+$ is the upper end-point of F (i.e. the smallest $z$ such that $F(z) = 1$). This means that the distribution of $M_n$ degenerates on $z_+$. The value of $z$ corresponding to FAPs near one will tend to infinity, thus causing the `instability' issue \citep{Coles01}.
\end{enumerate}

The Fisher--Tippett--Gnedenko theorem (also called the Extreme Value Theorem) lies at the foundation of EVT. It suggests that the maxima of a large sequence of i.i.d. univariate random variables after some standardization (see below) asymptotically follow well-defined distributions \citep{Fisher28, Gnedenko43}. The theorem's assumptions are very general, similar to those of the Central Limit Theorem for the asymptotic behavior of mean values. In both cases, the distribution $F$ of the variable does not need to be known; the theorem is valid for data drawn from almost all continuous distribution functions.

Consider again the maximum value $M_n$ of a sequence of $n$ i.i.d. random variables $x_1, x_2, ..., x_n$.  Start by standardizing a sequence of $n$ real-valued observations $M_n$ using sequences of constants \{$a_n > 0$\} and \{$b_n$\} to yield $M^*_n = \dfrac{M_n - a_n}{b_n}$. The Extreme Value Theorem states that, under broad conditions, the limiting distribution of $M^*_n$, $P(M^*_n \leq x)$, converges to distribution $G$ belonging to the Gumbel, Fr\'echet, or Weibull families. These families are conveniently combined into the Generalized Extreme Value (GEV) distribution \citep{Jenkinson55}, a three-parameter distribution with parameters $\mu$, $\sigma$, and $\xi$ denoting the location, scale, and shape:
\begin{equation}
    \begin{aligned}
    G(x) & = P(M^*_n \leq x)\\
         & = \textrm{exp} \Bigl\{-\left(1 + \xi \left( \dfrac{x - \mu}{\sigma} \right) \right)^{-1/\xi}\Bigl\}
    \end{aligned}
\end{equation}
where $1 + \xi \left(\dfrac{x - \mu}{\sigma}\right) > 0$ and $\sigma > 0$, $\mu \in \mathbb{R}$, and $\xi \in \mathbb{R}$. 
If the shape parameter $\xi = 0$, then $1 / \xi$ is undefined and the expression reduces to
\begin{equation}
G(x) = \textrm{exp} \Bigl\{ -\textrm{exp} \left( \dfrac{x - \mu}{\sigma} \right) \Bigl\}    
\end{equation}
by continuity; this is called the Gumbel distribution. The Fr\'echet and Weibull families correspond to $\xi > 0$ and $\xi < 0$, respectively. If $\xi > 0$ and $\xi \left( \dfrac{x - \mu}{\sigma} \right) \leq -1$, then G(x) = 0 and if $\xi < 0$ and $\xi \left(\dfrac{x - \mu}{\sigma} \right) \leq -1$, then G(x) = 1.  If one is interested in minimal rather than maximal values, the corresponding expressions for minima are obtained by replacing $x$ with $-x$. Probabilities for the GEV distributions are calculated with CRAN package \texttt{extRemes} \citep{Gilleland16}.

Two widely used approaches to evaluate the significance of extrema in EVT are the block-maxima and the peaks-over-threshold methods \citep{Coles01}. In the block-maxima approach, data is partitioned into blocks of a certain size, and the maximum from each block is used to generate a sample of extreme values on which the GEV model can be fit. The peaks-over-threshold selects extreme value samples by selecting values higher or lower than a chosen threshold, followed by declustering to try to achieve independence. The block-maxima method is a natural choice for periodograms because it accounts for short-range dependences in the periodogram and can be practically easier to use \citep{Ferreira15}.

An advantage of the EVT approach is that $n$, the number of independent values in the dataset, does not appear in the GEV distribution, as in equations (1)-(2). However, other difficulties seen with earlier FAP estimates remain. Harmonics of true periodicities with spectral peak strengths comparable to the true period will tend to overpopulate the extreme values and distort GEV probabilities. Heteroscedasticity and trends in the periodogram noise as a function of period violate the i.i.d. assumptions. This can be partially compensated by detrending and standardizing\footnote{The usual method of standardization is to subtract the mean and divide by the standard deviation. However, due to the non-Gaussianity of periodogram power values, we advocate a robust standardization that does not depend on normality; see equation (\ref{eqn:SNR}). \citet{Sulis17} provide a discussion of standardized periodograms for exoplanet detection from radial velocity observations.} the periodogram; that is, using local signal-to-noise ratios rather than periodogram power directly as advocated by \citet{Ofir14} and our treatment below (\S\ref{sec:per_snr}). 

Most importantly, the assumption that the dataset ${x_i}$ is i.i.d. (i.e., stationary white noise) will not apply to many astronomical periodograms. This assumption of the extreme value theorem can be relaxed under some circumstances; for example, the theorem is valid for dependent variables provided the dependence decays for increasingly separated variables when $n \to \infty$ \citep{Leadbetter88}. However, this condition is not met for oversampled periodograms.

Therefore, treating periodograms under EVT requires a computational approach in place of the asymptotic analytic formulae (3)-(4). The most widely used approach, nonparametric bootstrap resampling, is inadequate due to its assumptions of i.i.d. and the burden of computing periodograms with many trial periods for each bootstrap resample. \citet{Suveges14} propose a more efficient procedure involving bootstrapping partial periodograms over randomly chosen ranges of periods. This hybrid approach combining extreme value statistics and bootstrap resampling is favorably reviewed by \citet{VanderPlas18} and \citet{Koen21} for astronomical periodograms (see also \citealt{Cuypers12} for a more general take on extreme value distributions for peak significance). We adopt this approach here (\S\ref{sec:per_EVT}).

\section{Execution time comparison of BLS and TCF periodograms}
\label{sec:execution_time}
The computational execution time of the BLS and TCF calculations can be compared, and their scaling with the data set size can be established. The experiment was performed on a machine equipped with Intel(R) Xeon(R) CPU at 2.20GHz using a single CPU core. The same optimal frequency sampling is used for BLS and TCF.

We simulate light curves with Gaussian white noise and ten transits of a planet producing depth = 100 ppm transits, each with 2 hours duration. To increase the data points in the light curve, the period is progressively increased while all other parameters are kept fixed--the periods used are {0.5, 1, 3, 5, 10, 20, 40, 80} days. The number of points in the light curve ranges from a few hundred to $\sim$$10^4$. 

We use the \texttt{microbenchmark} CRAN package \citep{microbenchmark} to measure execution times. The execution time does not include preprocessing operations, such as ARIMA for TCF or Gaussian Processes regression before BLS; these occur much faster than the periodogram computation. For the relatively smaller periods, we use the median execution time across five different runs since we observed non-trivial differences across different runs. For the larger periods, we use the execution time from only one run.  

\begin{figure*}
    \centering
      \includegraphics[keepaspectratio,width=0.4\linewidth]{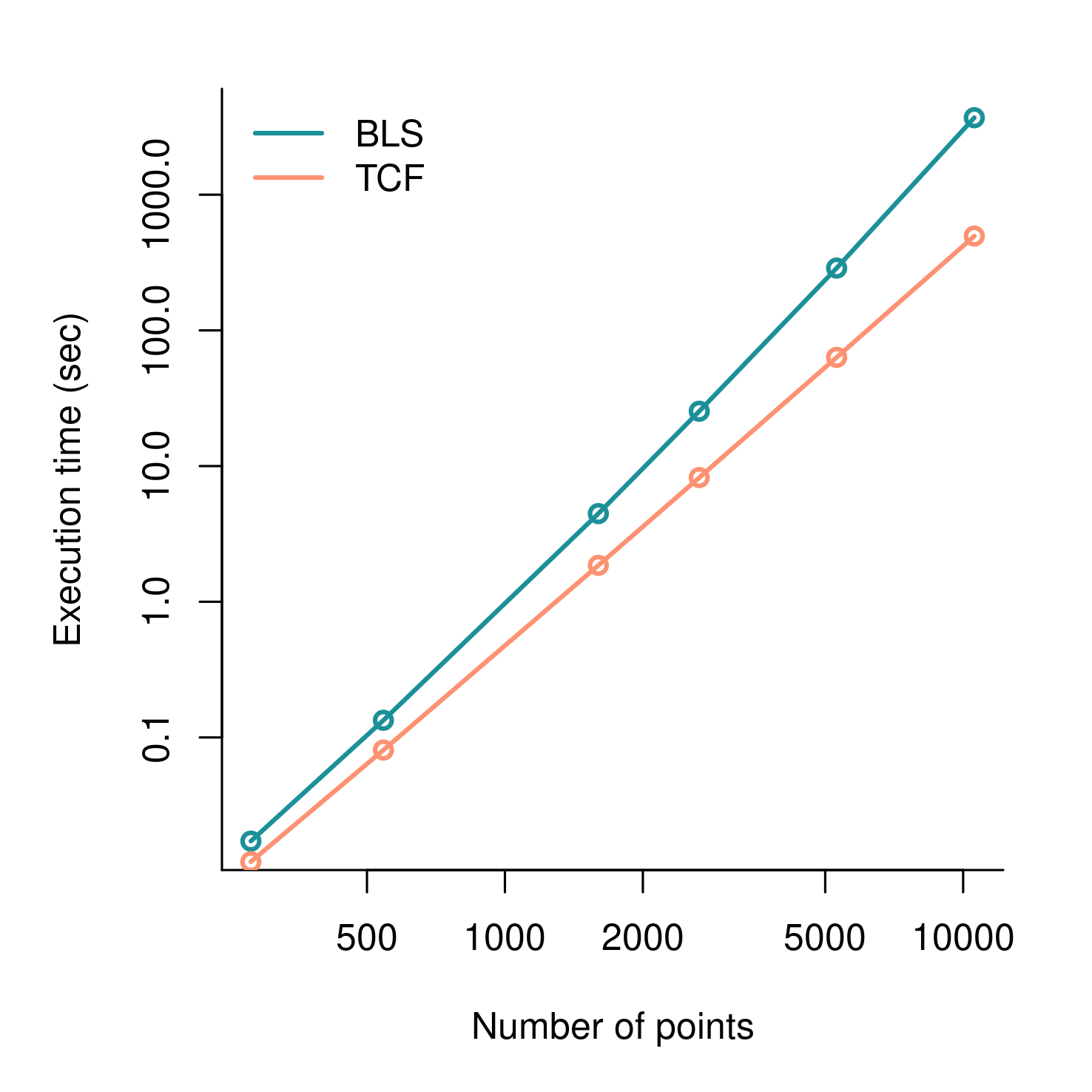}
    \caption{Execution time comparison of BLS and TCF as a function of the number of data points in the light curve.}
    \label{fig:time}
\end{figure*}

Figure ~\ref{fig:time} shows the execution time comparison of BLS and TCF. The figure illustrates that TCF is somewhat faster than the BLS algorithm and scales as $\mathcal{O} (N^{3})$, somewhat better than BLS with $\mathcal{O} (N^{3.7})$. We have not examined alternative, faster versions of the BLS algorithm such as SparseBLS \citep{Panahi21} and fBLS \citep{Shahaf22}.

\bibliography{periodograms}{}
\bibliographystyle{aasjournal}

\end{document}